%% file: susydefs.tex
\documentclass[12pt, utf8]{article}
\usepackage[utf8]{inputenc}

\usepackage[body={18cm, 23cm, centered}]{geometry}
\usepackage{amsmath,amssymb,amsthm,amscd,cite,comment,amsfonts,indentfirst,color,setspace,bbold,MnSymbol,arydshln}
\usepackage{mathtext,marvosym,textcomp}
\usepackage[makeroom]{cancel}
\usepackage{mathtools,slashed}
\usepackage{blkarray}
\usepackage[boxsize=1.2em]{ytableau}

\usepackage[debug,pageanchor=false]{hyperref}
\hypersetup{colorlinks=true,linktocpage,breaklinks,
	urlcolor=blue,
	linkcolor=red,
	citecolor=blue
}

\selectfont

\input{Defs.tex}

\numberwithin{equation}{section}

\begin{document}
\renewcommand{\refname}{\begin{center}References\end{center}}
	
\begin{titlepage}
		
	\vfill
	\begin{flushright}

	\end{flushright}
		
	\vfill
	
	\begin{center}
		\baselineskip=16pt
		{\Large \bf 
		SUSY and tri-vector deformations
		}
		\vskip 1cm
		    Alexander Kulyabin$^{a}$\footnote{\tt kulyabin.ak@phystech.edu},
			Edvard T. Musaev$^{a,b}$\footnote{\tt musaev.et@phystech.edu}, 	
		\vskip .3cm
		\begin{small}
			{\it 
				$^a$Moscow Institute of Physics and Technology,
			    Institutskii per. 9, Dolgoprudny, 141700, Russia,\\
				$^b$Kazan Federal University, Institute of Physics, Kremlevskaya 16a, Kazan, 420111, Russia\\
			}
		\end{small}
	\end{center}
		
	\vfill 
	\begin{center} 
		\textbf{Abstract}
	\end{center} 
	\begin{quote}
	     We analyze conditions for a tri-vector deformation of a supergravity background to preserve some supersymmetry. Working in the formalism of the SL(5) exceptional field theory, we present its supersymmetry transformations and introduce an additional USp(4) transformation to stay in the supergravity frame. This transformation acts on local indices and deforms BPS equations of exceptional field theory. The requirement for the deformation to vanish is the desired condition. The condition is shown to be consistent with previous results on bi-vector deformations.
	\end{quote} 
	\vfill
	\setcounter{footnote}{0}
\end{titlepage}
	
\clearpage
\setcounter{page}{2}
	
\tableofcontents

\section{Introduction}

Gauge/gravity duality, in its most general form, sets up a correspondence between solutions to supergravity equations of motion and gauge theories. The most well-understood example is the AdS/CFT correspondence that is an equivalence between gravitational degrees of freedom on the AdS${}_5\times \mathbb{S}^5$ background of Type IIB string theory and $\mc{N}=4$ $d=4$ super Yang--Mills theory, which is a superconformal gauge theory. The correspondence origins from the equivalence of two descriptions of the D3-brane: as a supergravity solution in terms of closed strings and as a world-volume theory in terms of open strings \cite{Maldacena:1997re}. Another well-known example is the correspondence between an AdS${}_4\times \SS^7$ background of 11-dimensional supergravity and the so-called ABJM theory \cite{Aharony:2008ug}, which is a $\mc{N}=6$ 3-dimensional Chern--Simons superconformal theory with gauge group SU(N), describing the world-volume theory of a stack of N M2-branes. These are particular examples of the more general observation \cite{Witten:1998qj} that the partition function of a theory with particular boundary data (given a boundary can be defined) can be rewritten as a partition function of a different theory:
\begin{equation}
    \label{eq:duality}
    \int_{\Phi(\dt M) = J} \DD \Phi e^{-i S[\Phi]} = e^{W[J]} = \int \DD \phi e^{-i \tilde{S}[\phi] - i \phi J}. 
\end{equation}
Here $J$ represents the values of the fields $\Phi$ on the boundary $\dt M$ of the $d+1$-dimensional space-time $M$, and $\phi$ denotes fields of the dual $d$-dimensional theory. The expression above might seem trivial as it is simply a sort of Laplace transformation of the effective action $W[J]$. The non-trivial part here is to determine whether the expression $\tilde{S}[\phi]$ can be interpreted as an action for a sensible theory. In addition to the examples above, which are pretty complicated and are based on string theory, one finds pairs of less involved theories:  (see \cite{Morozov:1998je} and references therein).

Due to its generality, the prescription \eqref{eq:duality} is not very suitable for searches of new pairs of dual theories, for which reason more algorithmic approaches become of particular interest. From the supergravity side, a powerful instrument is provided by solution-generating techniques based, in particular, on (non-abelian) T(U)-dualities and Yang--Baxter deformations. As an example, one may mention the Lunin--Maldacena solution to the supergravity equation that is obtained by a T-duality coordinate shift and T-duality (TsT) of the AdS$_5\times \SS^5$ background \cite{Lunin:2005jy}. This is known to be dual to the so-called $\beta$-deformation of Leigh--Strassler. The latter belongs to the most general superconformal three-parametric deformation of $\mc{N}=4$ $d=4$ SYM theory preserving $\mc{N}=1$ supersymmetry \cite{Leigh:1995ep}. The Lunin--Maldacena deformation is a particular case of the so-called Yang--Baxter bi-vector deformations that, for a given set of at least two Killing vectors $\{k_\a{}^m\}$ of the initial background $(g,b)$, can be written as (for the NS-NS sector) \cite{Araujo:2017jap,Araujo:2017jkb,Araujo:2017enj}
\begin{equation}
    \begin{aligned}
        (g'+b')^{-1} & = (g+b)^{-1} + \beta, \\
        \phi' & = \phi -\fr14 \log\fr{\det g}{\det g'}.
    \end{aligned} 
\end{equation}
Here $\beta^{mn} = r^{\a\beta}k_\a{}^mk_\beta{}^n$ is the bi-vector proportional to a constant matrix $r^{\a\beta}$. The $r$-matrix is required to satisfy the classical Yang--Baxter equation  and the so-called unimodularity constraint in order for the deformation to generate a solution
\begin{equation}
    \begin{aligned}
       r^{\beta_1 [\a_1}r^{\a_2|\beta_2|}f_{\beta_1\beta_2}{}^{\a_3]}&=0,\\
       r^{\a\beta}f_{\a\beta}{}^\g &=0.
    \end{aligned}
\end{equation}
Concluding that the formalism of (generalized) Yang--Baxter deformations serves as a useful tool for generating supergravity backgrounds, one becomes interested in interpreting the generated solution terms of dual field theories. A general rule can be implemented that a Yang--Baxter deformation on the field theory side is realized as a Drinfeld twist, corresponding to the given $r$-matrix \cite{vanTongeren:2015uha,Imeroni:2008cr}. At this step, it is important to determine whether a deformed supergravity solution preserves any of the supersymmetries of the initial one. For bi-vector deformation, determined by an r-matrix $r^{ab}$ satisfying a classical Yang--Baxter equation, this question was investigated in the works \cite{Orlando:2018kms,Orlando:2018qaq}, where a condition for deformation to preserve supersymmetry has been proposed. This is a non-linear differential condition on the bi-vector $\beta^{mn}$ that has first been derived explicitly for abelian deformations and then conjectured to be valid for non-abelian deformations. The conjecture has been successfully checked against various examples. 

When uplifted to 11-dimensional supergravity describing M-theory backgrounds, bi-vector deformations must naturally be generalized to tri-vector deformations, which has already been observed in \cite{Lunin:2005jy} for the abelian case. Since then, tri-vector deformations have been studied in a number of papers \cite{Ahn:2005vc,Gauntlett:2005jb,Berman:2007tf,Catal-Ozer:2009bvm,Hellerman:2012zf,Orlando:2018kms} and further from the point of view of the exceptional Sasaki--Einstein geometry in \cite{Ashmore:2018npi}. A description of tri-vector deformations in terms of symmetries of exceptional field theories, together with some explicit examples, was  first presented in \cite{Bakhmatov:2019dow,Bakhmatov:2020kul}. A more systematic approach was  developed in \cite{Gubarev:2020ydf} that allowed the fact that tri(six)-vector deformations always give solutions to supergravity equations given a generalization of the classical Yang--Baxter equation is satisfied was shown
. In this work, we continue this study and derive a condition for a tri-vector deformation to preserve supersymmetry. The main idea is to observe that when the fermionic sector of exceptional field theory (ExFT) is included, a tri-vector deformation, that is, an E$_{d(d)}$ transformation, must be accompanied by a local transformation $K$ that is an element of the maximal compact subgroup of E$_{d(d)}$. The reason is that a tri-vector deformation spoils the upper-triangular form of the generalized vielbein, thus moving the theory out of the supergravity frame. The latter is defined as the parametrization of the generalized vielbein and other fields of ExFT in terms of the supergravity fields, i.e., the metric and the 3-form field. To restore the upper-triangular form, one has to perform an additional transformation, which depends  on the tri-vector and background fields and acts non-trivially on fermions in general. The bosonic sector of exceptional field theory can be formulated purely in terms of a generalized metric that is a scalar under such transformation, and hence no additional rotation is needed. Hence, the criterion for deformation to preserve supersymmetry is that the transformed Killing spinor $\e' = K \cdot \e$ is again a Killing spinor, which eventually boils down to a condition on $K$ and hence on the tri-vector. 

For simplicity, we work in the SL(5) exceptional field theory and restrict ourselves to backgrounds of the form $M_4\times M_7$, where the deformation is performed on Killing vectors of the 4-dimensional manifold $M_4$. This is the same truncation as has been used in \cite{Bakhmatov:2020kul} to define deformations and earlier in \cite{Blair:2014zba} to generate a non-geometric U-duality partner of the M2-brane. To investigate BPS equations, we construct supersymmetry transformation rules for the theory following the same ideas as in \cite{Godazgar:2014nqa,Musaev:2014lna}, where supersymmetric versions of the E$_{7(7)}$ and E$_{6(6)}$ ExFT's have been constructed. Since there is no reason to believe that the general approach breaks for the SL(5) group, we do not go through the check of supersymmetry invariance of the full SL(5) ExFT action. Instead, we check commutation rules of supersymmetry transformation against the correct supersymmetry algebra and additionally compare to those of $D~=~7$ maximal supergravity, which is reproduced when fields of ExFT do not depend on the coordinates of the extended space.

The organization of the paper is as follows. In Section \ref{sec2}, we briefly introduce fields of exceptional field theory, supersymmetry transformations and generalized torsion constraints. In Section \ref{sec3}, we define the transformation of fermionic fields under tri-vector deformations and derive the condition where this preserves a Killing spinor. Finally, in Section \ref{sec4}, we apply the derived condition to some examples. First, we show that it reproduces the expected result upon reduction to 10 dimensions and bi-vector deformations. Second, we show that all tri-vector deformations of the M2-brane background that fit the SL(5) theory framework does not preserve supersymmetry. The same happens to be true for its near-horizon limit AdS$_4\times \mathbb{S}^7$, as we show that deformation commutes near the horizon limit
.

\section{Tri-Vector Deformations}\label{sec2}

In this section, we briefly review the SL(5) exceptional field theory and tri-vector deformation that belongs to the SL(5) U-duality group. Full construction of the bosonic sector can be found in \cite{Musaev:2015ces}; its truncation to only scalar fields is described by a DFT-like theory \cite{Berman:2010is,Berman:2011cg}, and the representative structure of the fermionic sector is the same as that of $D= 7$ supergravity \cite{Sezgin:1982gi}. In the construction of the supersymmetry transformation below, in this section, we follow the conventions of \cite{Samtleben:2005bp} for the local USp(4) indices and composite connections.

\subsection{Bosonic Sector of the SL(5) Theory}

Exceptional field theory is a reformulation of a supergravity covariant under U-duality group E$_{d(d)}$, where for $d=4,5$, the group is $\SL(5)$ and $\SO(5,5)$, respectively. The covariance is organized by decomposing fields of the 11-dimensional supergravity under the split $11=D+d$, collecting the obtained fields into irreducible representations of the global duality group as in \cite{Cremmer:1997ct}, and extending the $d$-dimensional space by an additional set of coordinates such that the corresponding derivatives $\dt_{\cM}$ fill an irreducible representation $\mc{R}_V$. Hence fields of the theory depend on a set of $D+\dim \mc{R}_V$ coordinates $(x^\m, \XX^{\cM})$ where the index $\mu$ labels  the so-called external coordinates $D$, and $\cM=1,\dots,\dim\mc{R}_V$ are the  coordinates of the extended space. The time direction can, in principle, belong to any of the two sets. For the SL(5) theory $\mc{R}_V = \mathbf{10}$,  it is convenient to label the extended coordinates as $\XX^{MN}$, where $M=1,\dots,5$, labels the fundamental {$\mathbf{5}$}  of SL(5). In this formalism, global SL(5) U-duality transformations are a particular coordinate transformation of the extended space. A general transformation can be written in the infinitesimal form as the so-called generalized Lie derivative \cite{Siegel:1993th,Siegel:1993xq,Berman:2012vc}: 
\begin{equation}
    \mL_\Lambda V^{M} = \fr12\Lambda^{KL}\dt_{KL} V^{M} - \fr{1}{4} \PP^{M}{}_{N}{}^{KL}{}_{PQ}\dt_{KL}\Lambda^{PQ} V^{N} + \fr12 \lambda_V \dt_{KL} \Lambda^{KL} V^M,
\end{equation}
where $\PP^M{}_N{}^{KL}{}_{PQ}$ is the projector on the adjoint representation of SL(5) (see Appendix \ref{sec:proj}). Explicitly one obtains 
\begin{equation}
    \mL_\Lambda V^{M} = \fr12\Lambda^{KL}\dt_{KL} V^{M} + \dt_{KL}\Lambda^{M K} V^{L} + \Big(\fr12 \lambda_V + \fr15\Big) \dt_{KL} \Lambda^{KL} V^M.
\end{equation}
Here $V^M$ is a generalized vector of weight $\lambda_V$ , and $\Lambda^{MN}$ is a parameter of the transformation. When the weight $\lambda_V = 1/10$, the above reproduces the usual expression for a generalized Lie derivative of the scalar sector of the SL(5) theory. Such defined transformations form a closed Lie algebra only if an additional section condition is imposed; that is
\begin{equation}
    \e^{MNKLP}\dt_{MN}\bullet \dt_{KL}\bullet =0,
\end{equation}
where bullets denote any combination of fields and theory derivatives. Basically, the condition restricts the dependence on the coordinates of the extended space \cite{Hohm:2010jy,Hohm:2013pua}. In what follows, we assume the section condition is solved such that all fields depend only on 4~coordinates out of 10, which corresponds to embedding of the full 11-dimensional supergravity.

The field content of the theory is the same as that of the $D~=~7$ maximal gauged supergravity; however, with fields depending on the coordinates of the extended space modulo the section constraint. The bosonic sector contains the metric, vector fields, 2-form and scalar matrix:
\begin{equation}
    \begin{aligned}
      & e_\m{}^\a, && A_\m{}^{MN}, && B_{\m\n M}, && \cV_M{}^{AB}.
    \end{aligned}
\end{equation}
Here small Greek letters from the beginning of the alphabet label local ``external'' directions,  those from the end of the alphabet label ``external'' coordinates, capital Latin indices $M,N,K,\dots=1,\dots,5$ label the $\mathbf{5}$ of SL(5), and capital Latin indices $A,B,C,\dots=1,\dots,4$ label the $\mathbf{4}$ of USp(4). In addition, there is a 3-form multiplet $C_{\m\n\rho}{}^M$ dual to the 2-form multiplet. These fields do not enter the equations of the theory. 

Scalar matrix $\mV_M{}^{[AB]}$ parametrizing the coset SL(5)/SO(5) contains a field scalar with respect to external coordinate transformations and contains the internal metrics $e_m{}^a$, 3-form field $C_{mnk}$, and a field $\phi$ proportional to a power of $\det e_\m{}^\a$. Index notations are self-evident. See that  the adjoint representation of SL(5) decomposes under its  GL(4) subgroup
\begin{equation}
    \begin{aligned}
    \mathbf{24} \to \mathbf{15}_0+\mathbf{1}_0+\mathbf{4}_{+5}+\mathbf{\bar{4}}_{-5},
    \end{aligned}
\end{equation}
where the subscript denotes weight with regard to  the GL(1) subgroup. Hence, the set of generators of SL(5) in GL(4) notations reads
\begin{equation}
    \bas \sl(5) = \{t_m{}^n, t_{mnk}, t^{mnk}\}.
\end{equation}
The coset element is then represented in the so-called triangular gauge as
\begin{equation}
    \mV = \exp[\phi \, t_{(0)}]\mV_4 \exp[C_{mnk}t^{mnk}],
\end{equation}
where $t_{(0)}$ is the GL(1) generator, and $\mV_4\in \SL(4)/\SO(4)$ corresponds to the standard non-linear realization of the 4-dimensional vielbein $e_m{}^a$. 

According to the prescription of \cite{Gubarev:2020ydf}, a tri-vector deformation $O_M{}^N$ is a (generalized) U-duality transformation generated by elements of negative weight
\begin{equation}
    \label{eq:3vector}
    O = \exp[\W^{mnk}t_{mnk}],
\end{equation}
and apparently does not preserve the triangular gauge. To restore that, one must complement the action of the deformation by a USp(4) transformation, which restores the triangular gauge. Upon the embedding $\SL(5) \hookleftarrow \USp(4)$, the adjoint representation decomposes as $    \mathbf{24} \to \mathbf{10} + \mathbf{14}$.
The space of irreducible representation $\mathbf{10}$ is spanned by symmetric tensors $T^{(AB)}$, while for elements of the space of $\mathbf{14}$, we have
\begin{equation}
    \begin{aligned}
    \mathbf{14}&\ni  T^{[AB]}{}_{[CD]}, && T^{AB}{}_{CB}=0, & T^{[AB]}{}_{[CD]}\Omega_{AB}=0=T^{[AB]}{}_{[CD]}\Omega^{CD},
    \end{aligned}
\end{equation}
where $\W_{AB}=-\W_{BA}$ is the invariant tensor of USp(4). In other words, starting with $T^{[AB]}$, such that $T^{AB}\W_{AB}=0$, parametrizing the $\mathbf{5}$, one observes that the decomposition of $\mathbf{5}\times \mathbf{5}$ contains precisely  $\mathbf{14}$. The conditions above remove the remaining irreducible representations in the decomposition.

Hence, the complement USp(4) transformation must be constructed using generators of SL(5), which remain in  $\mathbf{10}$ under the decomposition. Since $\mathfrak{usp}4=\mathfrak{so}(5)$ these are conveniently expressed in terms of SO(5) gamma-matrices $\G_{\bar{M}}{}^{A}{}_{ B}$:
\begin{equation}
    \begin{aligned}[]
        \{\G_{\bar{M}},\G_{\bar{M}}\} = 2\eta_{\bar{M}\bar{N}}, && \h=\mathrm{diag}[1,1,1,1,1],
    \end{aligned}
\end{equation}
where we introduce ``flat'' SO(5) indices $\bar{M},\bar{N},\bar{K},\dots=1,\dots,5$. 
For explicit calculations, we choose the following representation 
\begin{equation}
    \begin{aligned}
        \G_1&=\s_1\otimes 1, && \G_2= \s_2 \otimes 1, \\
        \G_3&=\s_3\otimes \s_1, && \G_4=\s_3\otimes \s_2, &&        \G_5=\s_3\otimes \s_3,
    \end{aligned}
\end{equation}
where $\s_1,\s_2,\s_3$ are the standard Pauli sigma matrices. We define gamma-matrices with all upper and all lower indices, such as the ones with indices raised and lowered by the invariant tensor $\W_{AB}=-\W_{BA}$ and its inverse $\W^{AB}$
, we derive the following symmetry properties
\begin{equation}
    \begin{aligned}
        &\mbox{antisymmetric}:&&\G_{\overline{M}}{}^{AB}, \G_{\overline{MNKL}}{}^{AB};\\
        &\mbox{symmetric}:&& \G_{\overline{MN}}{}^{AB},  \G_{\overline{MNK}}{}^{AB}.
    \end{aligned}
\end{equation}
Gamma-matrices $\G_{\bar{M}}{}^{AB}$ define the pseudo-real irreducible representation $\bf 5$ of USp(4) in their upper indices since they are traceless  $\G_{\bar{M}}{}^{AB}\W_{AB}=0$ and satisfy
\begin{equation}
    (\G_{\bar{M}}{}^{AB})^* = \W_{AC}\W_{BD}\G_{\bar{M}}{}^{CD}.
\end{equation}
Hence these can be used as coefficients relating the fundamental, irreducible representation of SL(5) to the pseudo-real irreducible representation $\bf 5 $ of USp(4) that allows the writing of
\begin{equation}
    \label{eq:54}
    \cV_M{}^{AB}=\fr{1}{2}\cV_M{}^{\bar{M}}\G_{\bar{M}}{}^{AB}.
\end{equation}
The prefactor is fixed by ensuring the usual definitions of the generalized metric:
\begin{equation}
    M_{MN}=\cV_M{}^{AB}\cV^{CD}\W_{AC}\W_{BD}=\cV_{M}{}^{\bar{M}}\cV_{N}{}^{\bar{N}}\eta_{\bar{M}\bar{N}}.
\end{equation}

\subsection{Fermions and Connections}

The fermionic sector of the theory contains the gravitino $\y_\m{}^A$ and dilatino $\chi^{ABC}$ fields, which have a hidden spinorial index. The spinors are symplectic Majorana, and the corresponding reality condition for a spinor $\psi^A$ reads
\begin{equation}
    \bar{\psi}_A^T = \W_{AB}C\y^B,
\end{equation}
where $C$ is the charge conjugation matrix, defined as
\begin{equation}
    \begin{aligned}
        & (\g^\m)^T = -C \g^\m C^{-1}, &&  C=C^T=-C^{-1}= -C^\dagger.
    \end{aligned}
\end{equation}
We define an SL(5) covariant derivative in the usual way as
\begin{equation}
    \nabla_{KL} V^M= \dt_{KL} V^M + \G_{KL, N}{}^M V^N +\fr53 \lambda_V, \G_{N[K,L]}{}^N V^M
\end{equation}
where generalized Christoffel symbols  are traceless $\G_{MN,K}{}^K = 0 $,  and each derivative $\dt_{MN}$ adds $-1/5$ to the weight of a tensor. The non-covariant part of the transformation of the Christoffel symbols is then
\begin{equation}
    \D_\Lambda \G_{KLN}{}^M = \dt_{KL}\dt_{NP}\Lambda^{MP} - \fr15 \dt_{KL}\dt_{PQ}\Lambda^{PQ}\delta^{M}_{N},
\end{equation}
which ensures the covariance of $\nabla_{MN}$ under a generalized Lie derivative. Denoting the weight of a spinor $\y^A$ in the $\bf 4$ of USp$(4)$ by $\lambda_\y$ we write for its covariant derivative
\begin{equation}
    \nabla_{MN} \y^A= \dt_{MN} \y^A - \fr14 \omega_{MN}{}^{\a\beta}\gamma_{\a\beta}\y^A - \cQ_{MN}{}_B{}^A \y^B +\fr53 \lambda_\y \Gamma_{K[M,N]}{}^K \y^A.
\end{equation}
The  generalized vielbein postulate
\begin{equation}
    \label{eq:postulate}
    0 = \nabla_{MN} \cV_K{}^{AB} = \dt_{MN} \cV_K{}^{AB} + 2\cQ_{MN}{}_C {}^{[A} \cV_K{}^{B]C} - \G_{MN,K}{}^L \cV_L{}^{AB}
\end{equation}
relates Christoffel  symbols to the composite connection   coefficients $\cQ_{MN}{}_B{}^A$. Note that the weight of the generalized vielbein is zero $\lambda_\cV =0$. In  turn, Christoffel  symbols can be fixed by imposing a vanishing torsion condition; that is 
\begin{equation}
    \label{eq:torsion}
    \mL^{\nabla}_\Lambda \cV_A^M - \mL_\Lambda^{\dt}\cV_A^M= \mT_{KL,N}{}^M \Lambda^{KL}\cV_A^N=0,
\end{equation}
where the superscript of the generalized Lie derivative denotes whether one uses the covariant or partial derivative. Using the fact that $\G_{MN,K}{}^L$ is traceless, we are able to write the generalized torsion as
\begin{equation}
    \mT_{KL,N}{}^M = \PP^M{}_N{}^P{}_Q \bigg[\fr12 \G_{KL,P}{}^Q + \G_{P[K,L]}{}^Q - \fr23 \G_{RP,[K}{}^R\delta_{L]}{}^Q\bigg].
\end{equation}
Explicitly, the vanishing torsion condition then takes the following form
\begin{equation}
    \label{eq:torsiongen}
    \fr32 \G_{[KLN]}{}^M - \G_{P[K,L}{}^P\delta_{N]}{}^M - \fr12 \G_{P(N,K)}{}^P\delta_L{}^M + \fr12 \G_{P(N,L)}{}^P\delta_K{}^M=0, 
\end{equation}
that implies that the torsion belongs to $\bf \overline{10}\times \bar{5} + 15$ of SL(5). Decomposing the condition into irreducible representations of USp(4) by explicit contraction of indices, we have
\begin{equation}
    \mT \in \bf 1 + 5 + 14 + 35'.
\end{equation}
On the other hand, Christoffel coefficients belong to the $\bf 10 \times 10$ of USp(4) that decomposes~as
\begin{equation}
   \bf 10 \times 10 \to 1 + 5 + 14+ 35' + 10 +35.
\end{equation}
We find that the vanishing torsion condition allows the fixing of the first four irreducible representations of generalized Christoffel indices in the decomposition above. The  $\bf 10$ is fixed by an additional constraint on the covariant derivative of the external vielbein
\begin{equation}
    \label{eq:gamma10}
    e_\a{}^\m \nabla_{KL} e_\mu{}^\a = 0 \quad \longrightarrow \quad \G_{MN} = -\fr37 e^{-1}\dt_{MN}e.
\end{equation}
The remaining $\bf 35$ is the undetermined part of the connection, which drops from all relevant expressions, such as the Lagrangian, BPS equations, etc. Having such a piece in Christoffel coefficients are the standard feature of generalized geometry  in double field theory \cite{Hohm:2011si} and in exceptional field theory \cite{Coimbra:2011ky,Coimbra:2012af,Cederwall:2013naa}.

Denoting $\mD_{MN} = \dt_{MN} + \cQ_{MN}$, it is convenient to express Christoffel symbols as 
\begin{equation}
    \G_{MN,K}{}^L= \cV_{AB}{}^L \mD_{MN}\cV_K{}^{AB}
\end{equation}
and substitute into \eqref{eq:torsiongen} to arrive at a condition on the composite connection coefficients. This can be rewritten in the following suggestive form
\begin{equation}
    \PP^{\bar{M}}{}_{\bar{N}}{}^{\bar{P}}{}_{\bar{Q}}\bigg[ \cV_{\bar{P}}{}^P\mD_{KL}\cV_P{}^{\bar{Q}} + 
    2 \cV_{\bar{P}}{}^P\mD_{P[K}\cV_{L]}{}^{\bar{Q}}
    \bigg]=0,
\end{equation}
where barred indices can be understood as symmetrized pairs of USp(4) indices $\bar{M}\leftrightarrow (AB)$. Since the expression in brackets above belongs to  algebra $\mathfrak{sl}(5)$, it can be decomposed into $\bf 10 + 14$ of USp(4), which gives the following conditions
\begin{equation}
    \begin{aligned}
        \cV_{CD}{}^P \mD_{KL}\cV_P{}^{AB} + 2 \cV_{CD}{}^P \mD_{P[K}\cV_{L]}{}^{AB} 
        = 0.
    \end{aligned}
\end{equation}
Defining, as usual, $Q_{MN A}{}^B = - \mV_{AC}{}^K \dt_{MN}\mV_{K}{}^{BC}$ the part of $\cV^{-1}\dt_{MN}\cV$ in $\mathbf{10}$, the above condition is explicitly solved as 
\begin{equation}
    \begin{aligned}
        \cQ_{M N, A}{}^B& = Q_{MN, A}{}^B + \cV_{MN}{}^{(CD)}\W^{BE}q_{CD,AE}, \\
        \mD_{MN}\cV_{K}{}^{A B} &= \cV_K{}^{C D}P_{MN, CD}{}^{A B} - 2 \cV_{MN}{}^{(EF)}\cV_K{}^{C[A}\W^{B]D}q_{EF,CD}.
    \end{aligned}
\end{equation}
{Here} $\cV_{MN}{}^{(CD)} = \cV_{[M}{}^{CA}\cV_{N]}{}^{DB}\W_{AB}$ and the tensor $q_{AB,CD} = q_{(AB),(CD)}$ contributing to the composite connection, in general, belongs to
\begin{equation}
    q_{AB,CD} \in \bf  10 \otimes 10 = 1 + 5 + 10 + 14 + 35 + 35'
\end{equation}
and should be constructed solely for $P_{AB,CDEF} = \cV_{AB}{}^{MN}P_{MN,CDEF} \in \bf 10\times 14$ to make the torsion vanish. Explicit calculations using the computer algebra system Cadabra \cite{Peeters:2007wn} show that the vanishing torsion condition contains only expressions of the type 
\begin{equation}
    \Omega^{EF}P_{E(A,B)FCD} \in \bf  35.
\end{equation}
{Moreover,} the $\bf 35'$ part of $q_{AB,CD}$ drops from the condition and is basically the undetermined part $u_{(ABCD)}$ of the connection. Hence, the only irreducible representation of $q_{ABCD}$ to be identified is the $\bf 35$. Hence, finally, we have
\begin{equation}
    \begin{aligned}
    q_{ABCD}=P^{\bf 35}_{CABD} + u_{ABCD},
    \end{aligned}
\end{equation}
where
\begin{equation}
         P^{\bf 35}_{A B C D }\in 
         \ytableausetup{centertableaux}
\begin{ytableau}
C & A & B \\
D 
\end{ytableau}\, .
\end{equation}
{In} tensor components, this reads
\begin{equation}
    \begin{aligned}
    P^{\bf 35}_{A B C D } &= \fr34 P_{(ABC)D} - \fr34 P_{(ABD)C}.
    \end{aligned}
\end{equation}
{This} completely determines the Christoffel symbols and hence the composite connection up to the undetermined part $u_{ABCD}$, which is of no interest.

\subsection{BPS Equations}

Supersymmetry transformations rules for fields of the SL(5) exceptional field theory have the following form
\begin{equation}
\begin{aligned}
\delta e_{\mu}^{\a}=&\ \frac{1}{2} \bar{\epsilon}_{A} \gamma^{\a} \psi_{\mu}^{A}, \\
 \delta A_{\mu}{}^{M N}=&-2\sqrt{2}V_{AB}{}^{[M } \mathcal{V}_{CD}{}^{N]} \Omega^{B D}\left(\frac{1}{2} \Omega^{A E} \bar{\epsilon}_{E} \psi_{\mu}{}^{C}+\frac{1}{4} \bar{\epsilon}_{e} \gamma_{\mu} \chi^{CAE}\right),\\
\delta \mathcal{V}_{M}{}^{AB}=&\ \frac{1}{4} \mathcal{V}_{M}{}^{CD}\left(\Omega_{E[C} \bar{\epsilon}_{D]}  \chi^{ABE}+\frac{1}{4} \Omega_{CD} \bar{\epsilon}_{E} \chi^{ABE}+\Omega_{CE} \Omega_{DF} \bar{\epsilon}_{G} \chi^{CF[A} \Omega^{B] G}\right.\\
&+\left.\frac{1}{4} \Omega_{CE} \Omega_{DF} \Omega^{AB} \bar{\epsilon}_{G} \chi^{CFG}\right), \\
\end{aligned}
\end{equation}
\begin{equation}
\begin{aligned}
\delta B_{\mu \nu M}=&\ 8\sqrt{2}\mathcal{V}_{M}{}^{AB}\left(-\Omega_{AC} \bar{\epsilon}_{B} \gamma_{[\mu} \psi_{\nu]}^{C}+\frac{1}{8} \Omega_{AC} \Omega_{BD} \bar{\epsilon}_{E} \gamma_{\mu \nu} \chi^{CDE}\right)\\
&+2 \sqrt{2}\epsilon_{M N P Q R} A_{[\mu}{}^{N P} \delta A_{\nu ]}{}^{Q R},\\
\Delta C_{\mu \nu \rho}{}^{M}=&\ \mathcal{V}_{AB}{}^{M}\left(-\frac{3}{8} \Omega^{AC} \bar{\epsilon}_{C} \gamma_{[\mu \nu} \psi_{\rho]}^{B}-\frac{1}{32} \bar{\epsilon}_{C} \gamma_{\mu \nu \rho} \chi^{ABC}\right),\\
\end{aligned}
\end{equation}
for bosonic fields and 
\begin{equation}
\begin{aligned}
\delta \psi_{\mu}{}^{A}&=\mathcal{D}_{\mu} \epsilon^{A}+\Omega_{CD } \mathcal{V}^{M AC} \mathcal{V}^{N BD} \Omega_{BF}\left( -\fr45 \nabla_{M N}^{+}\left(\gamma_{\mu} \epsilon^{F}\right)+\fr34 \gamma_{\mu} \nabla_{M N}^{+} \epsilon^{F}\right)\\
&-\fr{1}{15}\mathcal{V}_{BC}{}^{N} \mF_{\nu \rho \lambda N} \Omega^{AB}\left(\gamma^{\nu \rho \lambda}{ }_{\mu}+\frac{9}{2} \gamma^{\nu \rho} \delta_{\mu}^{\lambda}\right) \epsilon^{C},\\
\delta \chi^{ABC}&=2 \Omega^{CD} \mathcal{P}_{\mu DE}{}^{AB} \gamma^{\mu} \epsilon^{E} -8 \Omega_{ED} \mathcal{V}^{M CE} \mathcal{V}^{N [A  \mid D \mid} \nabla_{M N}^{+} \epsilon^{B]}\\
&+ \fr85  \left(\Omega^{AB} \delta_{G}^{C}-\Omega^{C[A} \delta_{G}^{B]}\right) \Omega_{DE} \Omega_{FH} \mathcal{V}^{M G F} \mathcal{V}^{N D H} \nabla_{M N}^{+} \epsilon^{E}  \\
& -\fr16\left(\Omega^{AD} \Omega^{BE} \mathcal{V}_{DE}{ }^{L} \mF_{\mu \nu \rho L} \gamma^{\mu \nu \rho} \epsilon^{C}-\frac{1}{5}\left(\Omega^{AB} \Omega^{CF}+4 \Omega^{C[A} \Omega^{B] F}\right) \mathcal{V}_{FE}{ }^{L} \mF_{\mu \nu \rho L} \gamma^{\mu \nu \rho} \epsilon^{E}\right)
\end{aligned}
\end{equation}
for fermionic fields, where 
\begin{equation}
    \begin{aligned}
    \mathcal{D}_{\mu} \epsilon^{A}&={D}_{\mu}\epsilon^{A}+ \frac{1}{4}\w_\m{}^{ab} \gamma_{ab}\epsilon^{A} + Q_{\mu}{}_B{}^A \epsilon^{B}, \\
        D_\m& = \dt_\m - \mc{L}_{A_\m}.
    \end{aligned}
\end{equation}
{In} addition, following the analogy with the E$_{6(6)}$ case, we define two shifted-covariant derivatives of spinors
\begin{equation}
    \begin{aligned}
        \nabla_{M N}^{\pm}\epsilon^{A}=&\ \partial_{M N}\epsilon^{A}+ \frac{1}{4}e^{\mu [\a} \partial_{M N}e_{\mu}^{\beta]} \gamma_{\a\beta}\epsilon^{A}   \pm  \frac{1}{4} \mF_{\mu \nu M N} e^{\mu \a} e^{\nu \beta} \gamma_{\a\beta}\epsilon^{A}\\
        &-\cQ_{MN}{}_B{}^A \epsilon^{B}   + \fr53 \lambda_\e \G_{K[M,N]}{}^K \e^A,\\
    \end{aligned}
\end{equation}
{The} derivative $\nabla_{MN}^+$ enters  the SUSY rules above, while $\nabla_{MN}^-$ might be necessary for writing the full supersymmetric action of the theory, as  was the case in \cite{Musaev:2014lna}. Note, however, that, e.g., in the supersymmetric E$_{7(7)}$ theory, one needs only one such derivative. We do not aim for the construction of the supersymmetric action, hence the derivatives $\nabla_{MN}^-$ will not be used here. Finally, the scalar current 1-form with components in the $\mathbf{14}$ of USp(4) is defined as usual as 
\begin{equation}
\mathcal{P}_{\mu}{ }^{ABCD}=D_{\mu} \mathcal{V}_{M}{ }^{[AB} \mathcal{V}^{CD] M},
\end{equation}

Although we do not construct a full invariant supersymmetric action for the theory, we check the above transformation rules by other means. First, the above  precisely reproduces the SUSY rules of the maximal {$D~=~7$} 
 supergravity when $\dt_M = 0$. Second, these form a Lie (super-)algebra together with other symmetries of the theory: $D~=~7$ Lorenz transformations, external diffeomorphisms, generalized diffeomorphisms, and gauge transformations. In Appendix \ref{app:susy}, we start with the most general form of supersymmetry transformations rendering $D~=~7$ SUSY rules upon $\dt_M=0$, which contain various numerical coefficients. Then, we fix these by requiring the correct commutation rules, which are supersymmetry transformations close to diffeomorphisms, both external and generalized, local SO(1,6) transformations, and gauge transformation. The only arbitrary coefficient left can be absorbed into a redefinition of, say, the field $C_{\m\n\rho}{}^M$.

\section{Deformation of Supersymmetry}\label{sec3}

Tri-vector-generalized Yang--Baxter deformation, as defined in \cite{Gubarev:2020ydf}, is a SL(5) transformation generated by elements of $\sl(5)$ with negative levels with regard to  the $\gl(4)$ decomposition, which preserves  generalized fluxes. The latter is constructed of a properly rescaled generalized vielbein. The rescaling is necessary to render the theory purely in terms of such fluxes and the external vielbein, i.e., truncate SL(5) ExFT to only the external gravity and internal scalar sector. Such theory describes only backgrounds of the form $M_4\times M_7$ with vanishing fields $A_\mu{}^{MN}$ and $B_{\m\n M}$. Among the backgrounds covered by the truncation are AdS vacuum solutions, which are of interest for holography applications and some M-theory and IIA brane solutions.

Explicitly,  rescaling is defined as
\begin{equation}
    \begin{aligned}
        e_\m{}^\a & = e^{-\f(y)}e^{\fr15}\bar{e}_\m{}^\a(x), \\
        \mV_M{}^{\bar{M}} & = e^{-\f(y)}e^{\fr15}V_M{}^{\bar{M}}(y), 
    \end{aligned}
\end{equation}
where $e = \det ||e_m{}^a(y)||$ denotes the determinant of the internal vielbein and is restricted to depend only on the coordinates $y^{m}$ parametrizing $M_4$. The same holds for $\f = \f(y)$ and $V_M{}^A=V_M{}^A(y)$, while $\bar{e}_\m{}^\a = \bar{e}_\m{}^\a(x)$ are functions of external coordinates $x^\m$ only. As it has been shown in \cite{Bakhmatov:2020kul}, this provides a consistent truncation to a subsector of the theory.  Explicitly, generalized vielbein for the truncated theory reads
\begin{equation}
    \label{eq:rescvielbein}
    V_M{}^{\bar{M}} = 
        e^{\fr\f2}\begin{bmatrix}
             e^{-\fr12}e_m{}^{a} & e^{\fr12}v^a \\
             0 & e^{\fr12}
        \end{bmatrix},
\end{equation}
where $v^a=e_m{}^a v^m$ and $v^m = 1/3! \varepsilon^{mnkl}C_{nkl}$. Tri-vector deformation is then 
\begin{equation}
    \label{eq:deform}
    \begin{aligned}
        V &\to V'= O\, V, \\
        O & = \exp[\W^{mnk}t_{mnk}] = 
            \begin{bmatrix}
                 \delta^m_n & e^{-1}W_n \\
                 0 & 1
            \end{bmatrix}.
    \end{aligned}
\end{equation}
{Here,} $W_m = 1/3! \varepsilon_{mnkl}\W^{nkl}$, and hence $O$ does not depend on the background fields.

\subsection{Local Deformation and Composite Connections}

The generalized vielbein \eqref{eq:rescvielbein} is in the upper-triangular form, which means a parametrization in terms of supergravity fields $e_\m{}^\a$, $e_m{}^a$ and $C_{mnk}$. The transformation \eqref{eq:deform} breaks this parametrization, introducing the left lower block. Note that the generalized metric defined as $m_{MN} = V_M{}^{\bar{M}}V_N{}^{\bar{N}}\delta_{\bar{M}\bar{N}}$ does not depend on the choice of parametrization and can always be understood as a matrix of the form
\begin{equation}
    m_{MN} =e^{\f} \begin{bmatrix}
         h^{-\fr{1}2} h_{mn} & - v_m \\
         -v_n & h^{\fr12}(1 + v_k v^k)
    \end{bmatrix}.
\end{equation}
{Here} $h_{mn} = e_m{}^a e_n{}^b h_{ab}$ and $h = \det ||h_{mn}||$. This allows to read-off transformations of the bosonic fields $\phi$, $h_{mn}$ and $C_{mnk}$ under tri-vector deformation. 

For the supersymmetric formulation of ExFT one, however, should use  vielbein rather than  metric, which makes it necessary to introduce an additional transformation that restores the upper-triangular frame. Since from the point of view of the Usp(4) subgroup breaking of the triangular gauge precisely introduces  the $\mathbf{10}$ part of the $\sl(5)$ algebra, to remove that, one should act by a transformation $K \in \USp(4)$ constructed exclusively of $\Omega^{mnk}$, space-time fields, and generators of USp(4). Now, the matrices $\G_{\bar{M}\bar{N}}{}^{A}{}_{B}$ are proportional to generators of SO(5) or equivalently USp(4), which allows us to write
\begin{equation}
    K_{(\mathbf{4})} = \exp[\alpha(W) W^a\G_{5a}],
\end{equation}
where we denote $W_a=\frac{1}{3!} \e_{abcd}\W^{bcd}$ (flat indices) and $\alpha(W)$ as some functions of $W^a$ and $V^a$ to be determined later. Explicit calculation shows\vspace{-2pt}
\begin{equation}
    K_{(\mathbf{4})}=\cos\big(\alpha(W) W\big) +   \fr{1}{W}\sin\big(\alpha(W) W\big)\, W^a \G_{5a} 
\end{equation}
with the obvious notation $W^2=W^aW_a$, note also $\det K = 1$. Function $\a(W)$ is determined by the condition that $K$ restores the supergravity frame. For that, we write the transformation in the representation $\bf 5$ of SO(5):
\begin{equation}
    K^{\bar{M}}{}_{\bar{N}} = \fr14 \G^{\bar{M}}{}_{AB}\G_{\bar{N}}{}^{CD}K^A{}_C K^B{}_D.
\end{equation}
{Explicitly,} in the component form, this reads
\begin{equation}
    \begin{aligned}
        K_{({\mathbf 5})}&=
        \begin{bmatrix}
             \delta^a{}_b - 2 \sin^2 (\a W) \dfr{W^a W_b}{W^2} &  \sin(2 \a W)\dfr{W^a}{W} \\
            -\sin(2 \a W)\dfr{W_b}{W} & \cos(2\a W).
        \end{bmatrix}\\
        &=
        \begin{bmatrix}
             \P^a{}_b & 0 \\
             0 & 0
        \end{bmatrix}+
        \begin{bmatrix}
             n^a n_b \cos(2\a W) & \sin (2\a W)n^a \\
             - \sin(2\a W) n_b & \cos(2\a W)
        \end{bmatrix},
    \end{aligned}
\end{equation}
where we define $n^a = W^a/W$ and the projector $\P^a{}_b = \delta^a{}_b - n^an_b$ on the hyperplane orthogonal to $W^a$.
Such defined $K_{(\mathbf{5})}$ restores the upper triangular gauge for $V$ if
\begin{equation}
    \label{eq:condition}
    \tan\left(2 \alpha(W) W \right) =  \frac{W}{1-W_av^a}.
\end{equation}
{Reduced} to only bi-vector deformations and $V_a=0$, the above gives the same condition as the one derived in \cite{Orlando:2018qaq}. Note the special case when $W_{m}v^m = 1$, where the above expression is not applicable. In this case, the condition for $K_{({\bf 5})}$ to restore the upper-triangular gauge is $W^2=0$ or $\cos (2\a W) = 0$. The former does not have non-trivial solutions in the Euclidean case, while the latter implies 
\begin{equation}
    2  \a(W) W = \fr \pi 2 + \pi n, \quad n \in\mathbb{Z}.
\end{equation}
{In} what follows, we assume $W_m v^m \neq 1$. Using the gamma matrix identity $\Gamma^{\bar{M}}{}_{CD}\Gamma_{\bar{M}}{}^{AB} = 4 \delta^{AB}{}_{CD} - \W^{AB}\W_{CD}$, one can rewrite the inverse relation between matrices in the $\bf 4$ and in the $\bf 5$:
\begin{equation}
    K^{[A}{}_C K^{B]}{}_D = \fr14 \Gamma^{\bar{M}}{}_{CD}\Gamma_{\bar{N}}{}^{AB}K^{\bar{N}}{}_{\bar{M}} - \fr14 \W^{AB}\W_{CD}.
\end{equation}

Given that (generalized) Yang--Baxter transformations are defined as such poly-vector deformations that preserve generalized fluxes, the vanishing torsion condition \eqref{eq:torsion} and the generalized vielbein postulate \eqref{eq:postulate} allow the relation of the composite USp(4) connection $\cQ_{MN,A}{}^B$ to components of the generalized flux. Indeed, the latter is defined as
\begin{equation}
    \label{eq:flux}
    \mL^\dt_{V_{AB,CD}}V_{EF}{}^M = \mF_{AB,CD,EF}{}^{GH} V_{GH}{}^M,
\end{equation} 
where the superscript $\dt$ again denotes that the generalized Lie derivative is written in terms of partial derivatives $\dt_{MN}$.  

Vielbein $V_{AB}{}^M$ is related to  vielbein $\cV_{AB}{}^M$ via the rescaling
\begin{equation}
    \cV_{AB}{}^M = \rho^{-1}V_{AB}{}^M,
\end{equation}
where $\rho=e^{-\fr\f2}e^{\fr1{10}}$ is a generalized scalar of weight $\lambda_\rho = 1/10$ and  vielbein $V_{AB}{}^M$ has weight $\lambda_V=\fr{1}{10}$. On the other hand, the vanishing torsion condition states that one can equivalently replace partial derivatives in generalized Lie derivative by $D=\dt+\G$, which for the rescaled vielbein reads
\begin{equation}
    D_{MN}V_{AB}{}^K = - 2 \cQ_{MN,[A}{}^C V_{B]C}{}^K + \rho^{-1}\nabla_{MN}\rho V_{AB}{}^K.
\end{equation}
{Moving} the composite connection term to the LHS we have
\begin{equation}
    0 = \nabla_{MN}\cV_{AB}{}^K = \nabla_{MN}(\rho^{-1}V_{AB}{}^K) = - \rho^{-2}\nabla_{MN}\rho \, V_{AB}{}^K + \rho^{-1}\nabla_{MN}V_{AB}{}^K.
\end{equation}
{Note} that we denote $\nabla$ as the fully covariant derivative, which includes all connections $\G$, $Q$, $\w$. This allows us to express the LHS of \eqref{eq:flux} in terms of the composite connection coefficients and $\dt_{MN}\rho$ and relate these to components of the generalized flux.  Rewriting $D_{MN}V_{AB}{}^K$ in terms of $Q_{MN A}{}^B$ we have
\begin{equation}
    \begin{aligned}
        \mF_{A B C D E F}{}^{G H}=&  - V_{A B CD}{}^{KL}\cQ_{K L E}{}^{A1}\Delta_{F A1}{}^{G H}  \\
        &+\cQ_{K L [C}{}^{A1}\Big(\Delta_{D] A1}{}^{G H}V_{AB E F}{}^{K L}- V_{D] A1}{}^{K}\Delta_{A B}{}^{G H}V_{E F}{}^{L}+ \frac{1}{2}V_{D] A1 A B}{}^{K L}\Delta_{E F}{}^{G H}\Big)\\
&- \cQ_{K L [A}{}^{A1}\Big(\Delta_{B] A1}{}^{G H}V_{C D E F}{}^{K L}+ V_{B] A1}{}^{K}\Delta_{C D}{}^{G H}V_{E F}{}^{L}- \frac{1}{2}V_{B] A1 C D}{}^{K L}\Delta_{E F}{}^{G H}\Big)\\
&+  \rho^{-1}\nabla_{K L}\rho\Big(V_{A B C D}{}^{KL}\Delta_{E F}{}^{G H}+ \Delta_{A B}{}^{G H} V_{C D E F}{}^{K L}-  \Delta_{C D}{}^{G H} V_{A B E F}{}^{K L}\Big),
\end{aligned}
\end{equation}
where we define $V_{ABCD}{}^{MN} = V_{AB}{}^{[M}V_{CD}{}^{N]}$. The left-hand side above is invariant under the generalized Yang--Baxter transformation; hence so is the right-hand side.

Let us now show that $\cQ_{ABC}{}^D = V_{AB}{}^{MN}\cQ_{MNC}{}^D$ contains  the same irreducible representations as  $\mF_{AB,CD}{}^{EF}$. Starting with the latter, we first notice that it belongs to the $\bf 10+15+\overline{40}$ of SL(5), which decomposes into 
\begin{equation}
    \mF_{AB,CD}{}^{EF} \in \bf 10 + 1 + 14 + 5 + 35'. 
\end{equation}
{For} the composite connection, we have
\begin{equation}
    \mQ_{AB,C}{}^D \in \bf 10\times 10 = 10 + 1 + 14 + 5 + 35' + 35.
\end{equation}
{The} last $\bf 35$ represented by a fully symmetric tensor of four indices, trivially drops from the LHS of \eqref{eq:flux}. This is the undetermined part of the connection, which does not enter BPS equations and will be obliviated from now on. Hence, we see that the irreducible representations inside the connection $\mQ_{AB,C}{}^D$ are precisely the same as the ones in the generalized flux. The only subtlety is with the $\bf 10$ part of the flux, which is the trombone, which contains an additional~term:
\begin{equation}
    \q_{(AB)} \propto \mF_{C(A,B)D,EF}{}^{EF}\W^{CD}=-\fr12\Big(Q_{C(A,B)}{}^C - 3 \rho^{-1}\nabla_{AB}\rho\Big) .
\end{equation}
{Imposing} the invariance of this combination, we define ``the invariant connection'' $\hat{\cQ}_{AB,C}{}^D$, which does not transform under generalized Yang--Baxter deformations
\begin{equation}
\hat{\cQ}_{AB,C}{}^D = \cQ_{AB,C}{}^D - \fr32\delta_{(A}{}^D \rho^{-1}\nabla_{B)C}\rho -\fr12 \W^{DE}\W_{C(A} \rho^{-1}\nabla_{B)E}\rho + \fr12  \rho^{-1}\nabla_{A B}\rho\delta_{C}{}^{D}
\end{equation}
{Furthermore}, under local USp(4) transformations, components of the generalized flux transform covariantly. 

This is an extremely important result for further narrative, as it allows us to take into account the complicated generalized Yang--Baxter equation by simply rewriting covariant derivatives in terms of the invariant composite connection. Note, however, that this has its own flaws as  we will be investigating what we call ``non-covariant'' parts of tri-vector transformation of BPS equations further. These are the differences between BPS equations written for the transformed Killing spinor on the transformed background and the initial BPS equations. Since we keep the connection $\hat{\mc{Q}}_{AB,C}{}^D$ invariant, the expression will be explicitly non-covariant with regard to  the internal 4-dimensional diffeomorphisms. However, this is only a consequence of the chosen approach, and the covariance is actually hidden, given the generalized Yang--Baxter equation and Killing vector conditions are taken into account. This simply follows from the fact that the initial BPS equation was covariant as is the tri-vector deformation and the new BPS equation on the new background. Hence the difference must also be covariant. Keeping  in mind that at some point, we simply restore explicit covariance by hand. This trick saves a huge amount of explicit calculations involving the generalized Yang--Baxter equation.

\subsection{Preserving Killing Spinors}

Consider now the BPS equations for the truncated theory, where we keep the initial (not rescaled) spinors 
\begin{equation}
\begin{aligned}
\delta \psi_{\mu}^{A}=&\ \mathcal{D}_{\mu} \epsilon^{A}-\fr45\Omega_{CD } \mathcal{V}^{M AC} \mathcal{V}^{N BD} \Omega_{BF}\left( \big(\nabla_{M N}\gamma_{\mu}\big) \epsilon^{F}+\fr25 \gamma_{\mu} \nabla_{M N} \epsilon^{F}\right),\\
\delta \chi^{ABC}=& -8\, \Omega_{ED} \mathcal{V}^{M CE} \mathcal{V}^{N [A  \mid D \mid} \nabla_{M N}\epsilon^{B]}+\fr{8}{5}  \left(\Omega^{AB} \delta_{G}^{C}-\Omega^{C[A} \delta_{G}^{B]}\right) \Omega_{DE} \Omega_{FH} \mathcal{V}^{M G F} \mathcal{V}^{N D H} \nabla_{M N}\epsilon^{E} 
\end{aligned}
\end{equation}
here
\begin{equation}
    \begin{aligned}
        \nabla_{M N}\epsilon^{A}&= \partial_{M N}\epsilon^{A}
        -\cQ_{MN}{}_B{}^A \epsilon^{B}   + \fr53 \lambda_\e \G_{K[M,N]}{}^K \e^A,\\
        \mD_\m \e^A & = \dt_\m \e^A + \fr14 \omega_\m{}^{\a\beta}\g_{\a\beta}\e^A.
    \end{aligned}
\end{equation}
{The} $\mathfrak{so}(1,6)$ spin-connection $\w_\m{}^{\a\beta}$ is defined by the vanishing torsion condition
\begin{equation}
    \dt_{[\m}e_{\n]}{}^\a + \w_{[\m\beta}{}^\a e_{\n]}{}^\beta =0,
\end{equation}
which gives
\begin{equation}
    \begin{aligned}
        \w_{\a\beta\g} &= \fr12 \Big(f_{\a\beta\g} - f_{\beta\g\a} + f_{\g\a\beta}\Big), \\
        f_{\beta\g}{}^\a & = -2 e_{[\beta}{}^\m e_{\g]}{}^\n \dt_{\m}e_\n{}^\a.
    \end{aligned}
\end{equation}
{Note} that the dilatino variation does not contain trace and antisymmetric parts:
\begin{equation}
    \begin{aligned}
        \delta \chi^{[ABC]}=0, && \delta \chi^{ABC}\W_{AB}=0.
    \end{aligned}
\end{equation}

Contracting the gravitino variation with $\g^\mu$ we have
\begin{equation}
    -\fr{25}{28}\bar{\g}^\m\mD_\m \e^A + \rho^{-1}\W^{AB}\Big(\nabla_{BC}\e^C + 5 \rho^{-1}\nabla_{BC}\rho\, \e^C\Big)=0.
\end{equation}
{Substituting} $\nabla_{AB}\e^B$ expressed from the above into the dilatino equation we have
\begin{equation}
    \begin{aligned}
        \delta \chi^{ABC}&=-4\, \nabla^{C[A}\epsilon^{B]}
        +4\,\left(\Omega^{AB} \W^{C D}-\Omega^{C[A} \W^{B] D}\right) \rho^{-1} \nabla_{DE}\rho\e^E\\
        &+\fr{5 \rho }{7} \left(\Omega^{AB} \W^{C D}-\Omega^{C[A} \W^{B] D}\right)\W_{DE}\bar{\g}^\mu \mD_{\m}\e^E=0, \\
    \end{aligned}
\end{equation}
where we use the following
\begin{equation}
    \begin{aligned}
        V_{MN}{}^{AB} & = \W_{C D}V_{M}{}^{AC}V_{N}{}^{BD},\\
        \dt_{M N} & = V_{MN}{}^{AB}\dt_{AB}, \\
        \dt_{AB} & = 2 V_{AB}{}^{MN}\dt_{MN},\\
        V_{MN}^{AB}V_{CD}{}^{MN} & = \fr12 \delta_{(C}{}^A\delta_{D)}{}^B.
    \end{aligned}
\end{equation}
{Here}, the second line is a definition of $\dt_{(AB)}$ and the third line follows from the identity in the fourth line.

Let us rewrite the derivative $\nabla^{C[A}\e^{B]}$ in terms of the connection $ \hat{\mQ}_{AB,C}{}^D$ that transforms covariantly under generalized Yang--Baxter deformation:
\begin{equation}
    \begin{aligned}
            \nabla^{C[A}\e^{B]} & = \hat{\nabla}^{C[A}\e^{B]} + \fr34 \rho^{-1}\nabla^{C[A}\rho \e^{B]} - \fr{3}{4}\Big(\W^{AB}\W^{CE} - \W^{C[A}\W^{B]E}\Big)\rho^{-1}\nabla_{ED}\rho \e^D\\
            \nabla_{AB}\e^B &= \hat{\nabla}_{AB}\e^B - 3 \rho^{-1}\nabla_{AB}\rho \e^B.
    \end{aligned}
\end{equation}
{Given} that the gravitino and dilatino equations become
\begin{equation}
    \begin{aligned}
        \delta \psi_{\mu}^{A}&= \mathcal{D}_{\mu} {\epsilon}^{A}-\fr{4}{25} \W^{AC} \rho^{-1}\bar{\g}_\m\left[  \hat\nabla_{CB}{\epsilon}^{B} + 2 \rho^{-1}\nabla_{AB}\rho\e^B\right]=0,\\
     \delta \chi^{ABC}&=4\,\bigg[\hat\nabla^{C[A} \e^{B]} + \fr34 \rho^{-1}\nabla^{C[A}\rho \, \e^{B]} -\fr{7}{4}\Big(\Omega^{AB} \W^{C D}-\Omega^{C[A} \W^{B] D}\Big) \W_{DE} \rho^{-1}\nabla_{DE}\rho\epsilon^{E}\bigg]  \\
     &\quad +\fr{5 \rho }{7} \left(\Omega^{AB} \W^{C D}-\Omega^{C[A} \W^{B] D}\right)\W_{DE}\bar{\g}^\mu \mD_{\m}\e^E=0 
     .
     \end{aligned}
\end{equation}
{Under} tri-vector deformations, we have the following transformation rules for the fields
\begin{equation}
    \begin{aligned}
        V'_M{}^{AB}& = K^A{}_CK^B{}_D V_N{}^{CD}O_M{}^N,\\
        \e^A{}' & = K^A{}_B \e^B,
    \end{aligned}
\end{equation}

For derivatives of the USp(4) spinor $\e^A$ this implies
\begin{equation}
    \label{eq:derrules}
    \begin{aligned}
       \hat\nabla'{}^{AB}\e'{}^C  =&\ K^A{}_EK^B{}_FK^C{}_G \hat\nabla^{EF}\e^G \\
        &+ 2  K^A{}_EK^B{}_F V^{EF MN}\dt_{MN}K^C{}_G \e^G
        + 4 K^A{}_EK^B{}_F V^{EF M L}\D_L{}^N \dt_{MN}K^C{}_G  \e^G \\
        & + 4 K^A{}_EK^B{}_F V^{EF M L}K^C{}_G  \Big[\D_L{}^N\dt_{MN}  \e^G + \fr53 \lambda_\e \D_L{}^N\G_{M N}  \e^G\Big],\\
    \end{aligned}
\end{equation}
where we have used the fact that $\hat\cQ_{AB,C}{}^D$ transforms covariantly and $\D_M{}^N= O_M{}^N - \delta_M{}^N$ has the only non-vanishing component $\D_m{}^5 = W_m$. The structure of the above expression is as follows. The first line is a covariant USp(4) transformation and will always vanish upon substitution into the BPS equations. The second line is the desired non-covariant part, which will define the supersymmetry preservation condition. The last line can be shown to vanish, given the Kosmann--Lie derivative of $\e^A$ vanishes. We show that we start with the last term of the last line and show that  $\D_{[L}{}^N \G_{M]N} = 3/5 \dt_{N[L}\D_{M]}{}^N$. For that, we recall the expression \eqref{eq:gamma10} for $\G_{MN}$ and consider the only non-vanishing components $\D_{[l}{}^N \G_{m]N}$:
\begin{equation}
    \begin{aligned}
        \D_{[l}{}^N \G_{m]N}& = \fr37 \tilde{W}_{[l} e^{-1}\dt_{m]}e =- \fr{3}{3!\, 7}\e_{pqr[l}\W^{pqr}e^{-1}\dt_{m]}e = -\fr{9}{3! \, 14} \e_{mlpq}\W^{pqr}e^{-1}\dt_r e\\
        &=-\fr{9}{3! \, 14} \e_{mlpq}e^{-1}\rho^{\a\beta\g}k_\a{}^p k_\beta{}^q k_\g{}^r\dt_r e = \fr{9}{3! \, 10} \e_{mlpq}\rho^{\a\beta\g}k_\a{}^p k_\beta{}^q \dt_r k_\g{}^r \\
        &=\fr{9}{3! \, 10} \dt_r \big(\e_{mlpq}\W^{pqr} \big)=-\fr{3}{5} \dt_{[m}\tilde{W}_{l]} = \fr35 \dt_{K[l}\D_{m]}{}^K.
    \end{aligned}
\end{equation}
{Here}, in the first line, we denote $\e_{mnkl}$ as the epsilon symbol; in the second line, we used $L_k e=0$ with the weight $\lambda[e_\m{}^\a]=1/5$, and in the last line, we used the unimodularity condition to move all Killing vectors under the derivative. To reshuffle indices, we used  $[mlpqr]\equiv 0$ in four dimensions. Hence  Equation \eqref{eq:derrules} takes the following form
\begin{equation}
    \label{eq:derrules1}
    \begin{aligned}
       \hat\nabla'{}^{AB}\e'{}^C  =&\ K^A{}_EK^B{}_FK^C{}_G \hat\nabla^{EF}\e^G \\
        &+ 2  K^A{}_EK^B{}_F V^{EF MN}\dt_{MN}K^C{}_G \e^G
        + 4 K^A{}_EK^B{}_F V^{EF M L}\D_L{}^N \dt_{MN}K^C{}_G  \e^G \\
        & + 4 K^A{}_EK^B{}_F V^{EF M L}K^C{}_G  \Big[\D_{L}{}^N \dt_{MN}\e^G - \lambda_\e \dt_{NL}\D_{L}{}^N \e^G\Big].
    \end{aligned}
\end{equation}
{Now} we notice that both the LHS of the above expression and the first line are covariant under local $\so(1,6)$ transformations and local coordinate shifts. Therefore,   the remaining terms must also be covariant
, although explicitly, the covariance is broken. As it was  advertised at the beginning of the section, this is the consequence of taking into account the generalized Yang--Baxter equation in the form of invariance of the connection $\hat{Q}_{AB,C}{}^D$. Hence, in principle, one may restore covariance explicitly, which we will not do at this step. 

Instead, we go further on the way of breaking explicit covariance by choosing a specific $\so(4)$ frame, where $L_k e_m{}^a=0$, which is certainly not true in general, even though the Killing vector condition $L_k g_{mn}=0$ holds. This allows us to show that in the chosen frame, the terms in brackets in \eqref{eq:derrules1} vanish, given the Kosmann--Lie derivative of $\e^A$ along Killing vectors $k_\a{}^m$ is zero. For that, we write 
\begin{equation}
    L_k \e^A = k^m D_m[\w] \e^A  + \fr14 \nabla_m[\G]k_n (\G^{mn})^A{}_B \e^B + \lambda_\e \dt_m k^m \e^A, 
\end{equation}
where $D_m[\w] = \dt_m + 1/4 \w_m{}^{ab}\G_{ab}$ is the standard $\mathfrak{so}(5)$ derivative and $\nabla_m[\G]$ is a derivative covariant with regard to  the standard Levi--Civita connection. Note the weight term. Using the Killing vector property, the first two terms above can be simplified as follows
\begin{equation}
    \begin{aligned}
        &\ k^m \dt_m \e^A  + \fr14 k^m\w_m{}^{ab}\G_{ab}{}^A{}_B\e^B + \fr14 e^m_a e^n_b\nabla_m[\G]k_n (\G^{ab})^A{}_B \e^B \\
        =&\ k^m \dt_m \e^A  + \fr14 k^m\w_m{}^{ab}\G_{ab}{}^A{}_B\e^B + \fr14 k^m e_{b n}\nabla_m[\G]e_a{}^n (\G^{ab})^A{}_B \e^B \\
        =&\ k^m \dt_m \e^A  + \fr14 k^m\w_m{}^{ab}\G_{ab}{}^A{}_B\e^B -  \fr14 k^m\w_m{}^{ab}\G_{ab}{}^A{}_B\e^B  = k^m \dt_m \e^A 
    \end{aligned}
\end{equation}
where in the second line, we used $L_k e_n{}^a = 0$ and the vanishing torsion condition, and in the third line, we used the vielbein postulate $\nabla[\G,\w]e_m{}^a=0$. In \cite{Orlando:2018kms} it was observed that for a spinor $\e^A$ to stay Killing after abelian T-duality, its Kosmann--Lie derivative along the corresponding isometry must be zero. Now the selected terms
\begin{equation}
    \label{eq:deveps}
    \D_{[M}{}^K \dt_{N]K}\e^A - \lambda_\e \dt_{K[N}\D_{M]}{}^K \e^A
\end{equation}
have only components $[MN]=[mn]$ given the section condition $\dt_{mn}=0$ and the structure of $\D_M{}^N$. These components  can be rewritten as
\begin{equation}
    \begin{aligned}   
        \fr14\e_{mnpq}\rho^{\a\beta\g}k_\a{}^pk_\beta{}^q 
        \Big[k_\g{}^r \dt_r \e^A + \lambda_\e \dt_r k_\g{}^r \e^A\Big]=0.
    \end{aligned} 
\end{equation}

Given all the above, the  non-covariant part of the transformation of $\hat \nabla^{AB}\e^C$ under a generalized Yang--Baxter deformation  simply becomes
\begin{equation}
    \begin{aligned}
        \D_\W\big(\hat \nabla^{AB}\e^C\big) =&\ 2  K^A{}_EK^B{}_F V^{EF MN}\dt_{MN}K^C{}_G \e^G.
    \end{aligned}
\end{equation}
{The} same analysis shows that the non-covariant transformation of $\nabla_{AB}\rho$ vanishes. Indeed, for the full transformations, we have 
\begin{equation}
    \nabla'_{AB}\rho  =K_A{}^C K_B{}^D \Big[\nabla_{CD}\rho + 4 V_{CD}{}^{ML}\D_L{}^{N}\nabla_{MN}\rho\Big],
\end{equation}
where the first term is the covariant transformation. For the only non-vanishing components of the second term, we write
\begin{equation}
    \begin{aligned}
        \D_{[m}{}^{K}\nabla_{n]K}\rho = -\fr14 
        \e_{mnpq}\rho^{\a\beta\g}k_\a{}^p k_\beta{}^q \Big[k_\g{}^r \dt_r \rho + \lambda_\rho \dt_r k_\g{}^r \rho\Big]\equiv 0.
    \end{aligned}
\end{equation}

As a result, we have the following transformations of the gravitino and dilatino supersymmetry variations 
\begin{equation}
    \begin{aligned}
        \delta \psi_{\mu}'{}^{A}&=K^A{}_B \delta\y_\m{}^B-\fr{4}{25} \W^{AB} \rho^{-1}\bar{\g}_\m\D_\W\left[  \hat\nabla_{BC} {\epsilon}^{C}\right]\\
      \delta \chi'{}^{ABC}&=K^A{}_D K^B{}_EK^C{}_F \delta\chi^{DEF}-4\D_\W\Big[\hat\nabla^{C[A} \e^{B]} \Big].
     \end{aligned}
\end{equation}
{Note} that the non-covariant transformation of gravitino is a trace of that of dilatino.  Hence, we conclude that for a spinor $\e^A$ to remain Killing under a tri-vector deformation, the non-covariant variations in the expressions above must vanish. This is a condition of the same sort as that imposed on generalized fluxes required  to transform them covariantly under SL(5) transformations to keep the equations of motion satisfied. Now we require the same for USp(4) transformations, which are necessary to restore the upper-triangular form of the generalized vielbein. Note that due to generalized Bianchi identities, this does not impose further constraints on $K^A{}_B$ from generalized fluxes.

We now calculate $\Delta_\W\hat{\nabla}^{AB}\hat{\e}^{C}$, which gives the main contribution to the supersymmetry preservation condition
\begin{equation}
    \begin{aligned}
        \Delta_\W\hat{\nabla}^{AB}\hat{\e}^{C}&=\sin(2\a W) 
        \fr1 4 e^{\f} K^{\bar{M}}{}_{\bar{K}}K^{\bar{N}}{}_{\bar{L}}
        V_{\bar{M}}{}^{M}V_{\bar{N}}{}^{L} \Big(2\delta_{L}^{N}+4 \Delta_{L}^{N}\Big) \G^{\bar{K} \bar{L}}{}^{AB}\partial_{MN}{K^{C}{}_D}{\e}^D \\
        &= W\Big({n}_{a} \sin(2\a W) {v}^{m} +  {\Pi}^{b}\,_{a} {e}_{b}\,^{m}  +  {e}_{b}\,^{m} {n}_{a} {n}^{b}  \cos(2\a W)\Big)  {\G}^{5 a}{}^{AB}{\nabla}_{m}{K^C{}_D}{\e}^D \overset{!}{=}0.\;
    \end{aligned}
\end{equation}
{Here} we replaced the partial derivative on $K$ with the ordinary $\gl(4)$ covariant derivative $\nabla_m=\dt_m +\G_m$ to restore the hidden covariance of the expression.

\subsubsection{Pure Metric Backgrounds: \texorpdfstring{$C_{mnk} = 0$}{C=0}}

For simplicity's  case, consider where the initial background has no gauge field, i.e., $v^m = 0$. Then $\a(W)$ depends only on $W^2 = W^a W_a$ and the derivative  $\dt_m K_{(\bf{4})}$ becomes particularly simple:
\begin{equation}
    \begin{aligned}
        \dt_m K_{(\bf{4})} &= \big(- \sin (\a W)  + \cos(\a W) (n\G)\big) (\a W)' \dt_m W + \sin (\a W) \dt_m n_a \G^{5a} \\
        &= K_{(\bf{4})} (n\G) (\a W)' \dt_m W + \sin (\a W) \dt_m n_a \G^{5a},
    \end{aligned}
\end{equation}
where prime denotes the derivative with regard to  $W$, and we denote $(n\G) \equiv n_a \G^{5a}$. Given  condition \eqref{eq:condition}, the derivative $(\a W)'$ can be rewritten as follows
\begin{equation}
    \label{eq:awder}
    \begin{aligned}
        (\a W)' = \fr12\cos^2(2\a W) = \fr{1}{2(1+W^2)}.
    \end{aligned}
\end{equation}

The antisymmetric pair of indices in the expression $\Delta_\W (\hat{\nabla}^{A[B}\hat{\e}^{C]})$ belongs, in general, to the $\bf 5 \oplus 1  $ of USp$(4)$. It is convenient to analyze these separately. Let us start with the singlet, which is
\begin{equation}   
    \label{eq:cond_singlet}
    \begin{aligned}
        &\Delta_\W (\hat{\nabla}^{A[B}\hat{\e}^{C]})\W_{BC}  = W\Big( {\Pi}^{b}\,_{a} {e}_{b}\,^{m}  +  {e}_{b}\,^{m} {n}_{a} {n}^{b}  \cos(2\a W)\Big)  {\G}^{5 a}{}^{A}{}_{B}{\nabla}_{m}{K^B{}_C}\hat{\e}^D   
    \end{aligned}
\end{equation}
{The} $\bf 5$ can be conveniently rewritten by contracting the above with $\G_{\bar{M} BC}$, which gives two sets of conditions, which are  $\bar{M}~=~5$ and $\bar{M}=a$. The former is the same as \eqref{eq:cond_singlet} multiplied by $\G^5$, while for the latter, we have
\begin{equation}
    \begin{aligned}
        &\Big[e_b{}^m - (1-\cos(2\a W))n^m n_b \Big]\G^{5b}\G_{a} \nabla_m K \hat{e} \\
    =&\ \Big[e_b{}^m - (1-\cos(2\a W))n^m n_b \Big]\big(2\G^5 \delta_a{}^b + \G_a \G^{5b}\big) \nabla_m K \hat{\e}.
    \end{aligned}
\end{equation}
{The} second term in the parentheses is proportional to \eqref{eq:cond_singlet} multiplied by $\G_a$ and hence vanishes, leaving us with the following condition
\begin{equation}
    \label{eq:thecond}
    e_a^n\Big[\delta_n{}^m - (1-\cos(2\a W))n^m n_n \Big]\G^5\nabla_m K \hat{\e}=0.
\end{equation}
{Multiplying} this by $\G^a$ we obtain precisely \eqref{eq:cond_singlet}; hence \eqref{eq:thecond} is the only condition for a spinor to remain Killing. Finally, the determinant of the matrix $\mc{O}^m{}_n = \delta_n{}^m - (1-\cos(2\a W))n^m n_n $ is equal to $\cos(2\a W)$, which does not vanish for any finite value of $W$. Hence, it does not have zero eigenvalues, which implies
\begin{equation}
   \nabla_m K_{({\bf 4})} \hat{\e}=0.
\end{equation}
{Hence}, the condition for a spinor to remain Killing is that it belongs to the kernel of  map $\nabla_m K$, where $K$ is the local tri-vector transformation, which restores the supergravity frame. 

The obtained condition can be further rewritten in a more convenient form in terms of deformation parameters $W_m$. For that, we first observe that since $\det K_{({\bf 4})}=1$, we can safely multiply the above by another copy of $K_{({\bf 4})}$ to have $\dt_m (K_{({\bf 4})}K_{({\bf 4})})\hat{\e}=0$. Given  condition \eqref{eq:condition}, the derivative can be easily calculated as follows
\begin{equation}
    \begin{aligned}
        \nabla_m (K_{({\bf 4})}K_{({\bf 4})}) &= \nabla_m \Big(\cos(2\a W)\big(1+ (W\G)\big)\Big)\\
        &=-\sin(2\a W) 2 (\a W)' (1+(W\G))\dt_m W + \cos(2\a W) \nabla_m W_n \G^{5n}\\
        &= -2\cos(2\a W) (\a W)'\Big[(1+(W\G))W\dt_m W - (1+W^2)\nabla_m W_n \G^{5n} \Big]\\
        &= -2 \cos(2\a W)(\a W)'(1+ (W\G))\Big[W\dt_m W - (1-(W \G))\nabla_m W_n \G^{5n}\Big]
    \end{aligned}
\end{equation}
where $(W\G) = W_a \G^{5a}$. Here, in the third line, we used  relation \eqref{eq:awder} for the derivative $(\a W)'$ and in the last line simply factored out $1+(W\G)$. Now, we notice that neither of the terms outside the brackets in the last line vanishes for finite values of $W$. Hence, we are left with the condition 
\begin{equation}
    \label{eq:cond1}
    \Big[W\dt_m W - (1-(W \G))\nabla_m W_n \G^{5n}\Big]\hat{\e}=0.
\end{equation}
{Finally}, writing $W \dt_m W = W^k \nabla_m W_k$ and expanding the parentheses we obtain the final result
\begin{equation}
    \label{eq:susycondzero0}
    \Big[\nabla_m W_n \G^{5n} + W_k \nabla_m W_n \G^{kn}\Big] \hat{\e}=0.
\end{equation}

\subsubsection{Backgrounds with Non-Vanishing 3-Form}

To generalize the above backgrounds with non-vanishing $v^m$ it is enough to make the following two observations. First, the matrix 
\begin{equation} 
    \mc{O}^m{}_n = \delta_n{}^m - (1-\cos(2\a W))n^m n_n  + \sin(2\a W)n_n v^m
\end{equation}
has determinant $\det \mc{O} = \cos(2\a W)\big(1-(v\cdot n)\big)^{-1}$, where  $(n \cdot v) = n_m v^m$,  which is never zero, as discussed above. Hence, $ \mc{O}^m{}_n$ is non-degenerate, and the condition for a spinor to remain Killing still has the form
\begin{equation}
   \nabla_m K_{({\bf 4})} \hat{\e}=0,
\end{equation}
with dependence on $v^m$ hidden in $K_{(\bf 4)}$. The second observation is that $\a W$ is actually a function of a new single variable $\w$, which is a combination of $W$ and $v^m$:
\begin{equation}
    \tan (2\a W) = \fr{W}{1-W_m v^m} = \fr{1}{W^{-1} - (n\cdot v)}=:\w.
\end{equation}
{Hence}, all steps of the previous case can be repeated with $(2\a W)'$ now meaning a derivative with regard to  $\w$. In particular, we have
\begin{equation}
    (2\a W)_\w' = \cos^2(2\a W) = \fr{1}{1+\w^2}.
\end{equation}
{Condition} \eqref{eq:cond1} now becomes
\begin{equation}
    \label{eq:cond2}
    \Big[\w\dt_m \w - \big(1-(n \G) \w\big)\dt_m \big(n_a \G^{5a}\w\big)\Big]\hat{\e}=0. 
\end{equation}

The form of the condition above suggests the definition $\w_m\equiv \w n_m$, which allows us to repeat all the steps from the previous case with $W_m \to \w_m$ to arrive at
\begin{equation}
    \label{eq:susycondzero}
    \begin{aligned}
        0&=\Big[\nabla_m \w_n \G^{5n} + \w_k \nabla_m \w_n \G^{kn}\Big] \hat{\e},\\
        \w_m& = \fr{W_m}{1- W_nv^n}.
    \end{aligned}
\end{equation}
{This} is the final equation in the form most convenient for direct calculations. Note that both $\nabla_m W_n$ and $\nabla_m \w_n$ are symmetric given the unimodularity constraint. Indeed, we~write 
\begin{equation}
    \nabla_m \w_n = \fr{1}{1-Wv}\bigg[\nabla_m W_n + \fr{W_m W_n}{1- Wv}\nabla_k v^k\bigg],
\end{equation}
where $\nabla_m v^m = \fr1{4!} \ve^{mnkl}F_{mnkl}$, which implies $\nabla_{[m}\w_{n]}=0$.

\section{Examples}\label{sec4}

Equation \eqref{eq:susycondzero} is the condition for a spinor $\e^A$ to remain Killing under a tri-vector deformation parametrized by $W_m=1/3!\e_{mnkl}\W^{nkl}$. This is a differential condition on the deformation tensor $W_m$ such that an operator can be constructed, which projects the spinor $\e$ to zero. Spinors belonging to the kernel of this operator remain Killing. As we will see below, for the considered setup, the condition is very restrictive, and for the most interesting and accessible cases, such as the AdS$_4\times \SS^7$ background, the kernel contains only zero spinors.

\subsection{Reduction to Ten Dimensions}

Let us first compare the condition obtained above for pure metric backgrounds to the condition of \cite{Orlando:2018qaq} for a bi-vector deformation to preserve the Killing vector. For that, we assume the unimodularity of the corresponding bi-vector deformations and keep only  component $W_{\bar{m}}$ with $\bar{m}=1,2,3$ labeling three directions of the 10~=~7 + 3 decomposition. In this case, we observe that the first term of \eqref{eq:susycondzero} reproduces precisely the same condition as that of~\cite{Orlando:2018qaq}, given no R-R fields are present. The quadratic term can be shown to vanish, for which we consider $\nabla_{\bar{m}}W_{[\bar{k}}W_{\bar{n}]}$. Contracting this with $\e^{\bar{n}\bar{k}\bar{l}}T_{\bar{l}}$ where $T_{\bar{l}}$ is arbitrary, we have
\begin{equation}
    \begin{aligned}
        W_{\bar{n}}  \nabla_{\bar{m}}W_{\bar{k}} \e^{\bar{n}\bar{k}\bar{l}}T_{\bar{l}} & = \fr{1}{3!3!}\e_{\bar{n}\bar{p}\bar{q}}\e_{\bar{k}\bar{r}\bar{s}}\beta^{\bar{p}\bar{q}}\nabla_{\bar{m}}\beta^{\bar{r}\bar{s}}\e^{\bar{n}\bar{k}\bar{l}}T_{\bar{l}} = -\fr{1}{3!3!}\e_{\bar{p}\bar{n}\bar{k}}\e_{\bar{q}\bar{r}\bar{s}}\beta^{\bar{p}\bar{q}}\nabla_{\bar{m}}\beta^{\bar{r}\bar{s}}\e^{\bar{n}\bar{k}\bar{l}}T_{\bar{l}}\\
        &=-\fr{2}{3!3!}\e_{\bar{q}\bar{r}\bar{s}}\beta^{\bar{p}\bar{q}}\nabla_{\bar{m}}\beta^{\bar{r}\bar{s}}T_{\bar{p}} =\fr{2}{3!3!}\e_{\bar{m}\bar{r}\bar{s}}\beta^{\bar{p}\bar{q}}\nabla_{\bar{q}}\beta^{\bar{r}\bar{s}}T_{\bar{p}} \\
        &= 
        \fr{2}{3!}\e_{\bar{m}[\bar{r}\bar{s}}\beta^{\bar{p}\bar{q}}\nabla_{\bar{q}}\beta^{\bar{r}\bar{s}}T_{\bar{p}]} \equiv 0,
    \end{aligned}
\end{equation}
where in the first line, we used antisymmetrization in four indices $[\bar{p}\bar{n}\bar{k}\bar{q}]=0$, and in the second line, we first used antisymmetrization in $[\bar{q}\bar{r}\bar{s}\bar{m}]=0$ together with the unimodularity constraint $\nabla_{\bar{m}}\beta^{\bar{m}\bar{n}}=0$. 

Hence, we conclude that our condition for a tri-vector deformation of 11D backgrounds of a certain form to preserve a Killing spinor agrees with the same for general bi-vector deformations of 10D backgrounds of \cite{Orlando:2018qaq}. In principle, the approach we develop here allows us to drop all the restrictions and derive a generalization of the condition valid for any 11D backgrounds.

\subsection{Membranes and Near-Horizon Geometry}

Let us first illustrate the method on the example of the M2-brane solution, which is a 1/2BPS background, i.e., preserves 16 spinors. For $N$ M2-branes, the background metric and gauge field can be written in the following form:
\begin{equation}
    \label{eq:m2bg}
\begin{aligned}
d s^2&=H^{-2 / 3}\left(-d t^{2}+d x^{2}+d y^{2}\right)+H^{1 / 3}\left(d r^{2}+r^{2} d \Omega_{7}^{2}\right), \\
C_{t x y}&=-H^{-1}, \,\, H=1+\frac{L^{6}}{r^{6}}.
\end{aligned}
\end{equation}
{Here} $L=2^{5 / 6} \pi^{2 / 6} N^{1 / 6} l_{p}$ with $l_p$ denoting the Planck length. We choose the longitudinal coordinates $x^0,x^1,x^2$ and the radial coordinate $r$ to be  internal. Hence the fields for the truncated SL(5) ExFT read:
\begin{equation}
    \begin{aligned}
        h_{mn} & = \diag\Big[H^{-\fr23},H^{-\fr23},H^{-\fr23},H^{\fr13}\Big], \\
        V^m  &= \Big[0,0,0,H^{-\fr16}\Big],\\
        e^{-\f} & = r\, H^{\fr16},
    \end{aligned}
\end{equation}
and  metric $\bar{g}_{\m\n} $ invariant under tri-vector transformations is that of the transverse $\SS^7$. The relevant isometry group is $\SO(1,2)\ltimes \RR^{3}$, which is the Poincare symmetry group of the world volume. Denoting generators $P_\a$ and $M_{\a\beta}$ with $\a=0,1,2$ we, in principle, can construct deformations with terms proportional to the coordinates $x^\a$ in zero,  first, second, and third powers. However, the only unimodular tri-vector deformation here is $\W^{012}=-\rho$, i.e., the abelian PPP deformation. In this case
\begin{equation}
    W_m = \Big[0,0,0, \rho H^{-\fr56}\Big].
\end{equation}
The condition  \eqref{eq:susycondzero} simply boils down to the system of equations
\begin{equation}
    \label{eq:system}
    \begin{aligned}
        \Big(\rho - (\rho + H) \G^{4}\Big) \e&=0, \\
        \G^{4}\e&=0,
    \end{aligned}
\end{equation}
which does not have non-trivial solutions. Hence, we conclude that the M2-brane background does not have tri-vector deformations that preserve SUSY within the SL(5) setup.

A similar conclusion can be made for the AdS${}_4\times \SS^7$ solution, which is the near-horizon limit of the previous background. To see that, we 
choose a new coordinate $u$ as
\begin{equation}
r=\frac{N^{1 / 4} l^{3 / 2}}{\sqrt{u}}
\end{equation}
and rescale $(x,y,t) \to  \frac{1}{2}(x,y,t)$, $l=N^{-1/6}L$ to rewrite  solution \eqref{eq:m2bg} as
\begin{equation}
\begin{aligned}
d s^2=&\ \frac{1}{4}\left(1+N^{-1 / 2} l^{-3} u^{3}\right)^{-2 / 3}\left(-d t^{2}+d x^{2}+d y^{2}\right)\\
&+\left(1+N^{-1 / 2} l^{-3} u^{3}\right)^{1 / 3} l^{3} N^{1 / 2}\left(\frac{1}{4 u^{3}} d u^{2}+\frac{1}{u} d \Omega_{7}^{2}\right), \\
C_{t x y}=&-\frac{1}{8}\left(1+N^{-1 / 2} l^{-3} u^{3}\right)^{-1}
\end{aligned}
\end{equation}
{The} near-horizon limit, giving the $AdS_4 \times S^7$ solution,  can then be performed as follows:
\begin{equation} 
\begin{aligned}
d s_{(h)}^2 &\equiv \lim _{N \rightarrow 0} \frac{d s^{2}}{N^{1 / 3}}=\frac{l^{2}}{4 u^{2}}\left(-d t^{2}+d x^{2}+d y^{2}+d u^{2}\right)+l^{2} d \Omega_{7}^{2}, \\
C_{(h) t x y} &\equiv \lim _{N \rightarrow 0} \frac{C_{t x y}}{N^{1 / 2}}=-\frac{1}{8} l^{3} u^{-3}.
\end{aligned}
\end{equation}
{Interestingly} enough,  as for the bi-vector case analyzed in \cite{Borsato:2018idb}, taking the near-horizon limit commutes with tri-vector deformations, meaning that the latter descends to world-volume theories. Let us demonstrate that using explicit examples of two types of deformations: PPP and PPM. 

{We start with $P \wedge P \wedge P$ deformation with $\Omega$-shift given by $t,x,y$ coordinate \mbox{translations}:}
\begin{equation} 
\Omega= 4 \eta \partial_{t} \wedge \partial_{x} \wedge \partial_{y}
\end{equation}
{Using} explicit formulas for tri-vector deformations, for the deformed background, we get 
\begin{equation} 
\begin{aligned}
ds^2& =\frac{1}{4}(1+\eta(\left.\left.+N^{-1 / 2} l^{-3} u^{3}\right)^{-1}\right)^{-2 / 3}\left(1+N^{-1 / 2} l^{-3} u^{3}\right)^{-2 / 3}\left(-d t^{2}+d x^{2}+d y^{2}\right) \\
&+\left(1+\eta\left(1+N^{-1 / 2} l^{-3} u^{3}\right)^{-1}\right)^{1 / 3} \frac{1}{4} \frac{l^{3} N^{1 / 2}}{u^{3}}\left(1+N^{-1 / 2} l^{-3} u^{3}\right)^{1 / 3} d u^{2} \\
&+\left(1+\eta\left(1+N^{-1 / 2} l^{-3} u^{3}\right)^{-1}\right)^{1 / 3}\left(1+N^{-1 / 2} l^{-3} u^{3}\right)^{1 / 3} \frac{l^{3} N^{1 / 2}}{u} d \Omega_{7}^{2}, \\
C^{ t x y} &=-\frac{1}{8} \left(1+\eta\left(1+N^{-1 / 2} l^{-3} u^{3}\right)^{-1}\right) \left(1+N^{-1 / 2} l^{-3} u^{3}\right)^{-1}.
\end{aligned}
\end{equation}
{To} go to the near-horizon area, we write $\eta=\hat{\eta} N^{-1 / 2}$ and keep $\hat{\eta}$ fixed in the limit, which gives:
\begin{equation} 
\begin{aligned}
d s_{(h)}^2&=\frac{l^{2}}{4 u^{2}}\left(1+\hat{\eta} \frac{l^{3}}{u^{3}}\right)^{-2 / 3}\left(-d t^{2}+d x^{2}+d y^{2}\right)+\left(1+\hat{\eta} \frac{l^{3}}{u^{3}}\right)^{1 / 3}\left(\frac{l^{2}}{4 u^{2}} d u^{2}+d \Omega_{7}^{2}\right), \\
C_{(h) t x y}& =-\frac{1}{8}\left(1+\hat{\eta} \frac{l^{3}}{u^{3}}\right)^{-1}.
\end{aligned}
\end{equation}
{This}  exactly reproduces $P \wedge P \wedge P$ deformation of the AdS${}_4 \times \SS^7$ solution of \cite{Bakhmatov:2020kul}.
 
 For $P \wedge P \wedge M $ deformation, we have
 \begin{equation} 
 \Omega=4 \rho_{\dot{\alpha}} x^{\dot{\alpha}} \partial_{t} \wedge \partial_{x} \wedge \partial_{y},
 \end{equation}
 which  for the deformed background gives:
\begin{equation} 
\begin{aligned} 
d s^2&=\frac{1}{4}\left(1+\rho_{\dot{\alpha}} x^{\dot{\alpha}}\left(1+N^{-1 / 2} l^{-3} u^{3}\right)^{-1}\right)^{-2 / 3}\left(1+N^{-1 / 2} l^{-3} u^{3}\right)^{-2 / 3}\left(-d t^{2}+d x^{2}+d y^{2}\right) \\
&+\left(1+\rho_{\dot{\alpha}} x^{\dot{\alpha}}\left(1+N^{-1 / 2} l^{-3} u^{3}\right)^{-1}\right)^{1 / 3} \frac{1}{4} \frac{l^{3} N^{1 / 2}}{u^{3}}\left(1+N^{-1 / 2} l^{-3} u^{3}\right) d u^{2} \\
&+\left(1+\rho_{\dot{\alpha}} x^{\dot{\alpha}}\left(1+N^{-1 / 2} l^{-3} u^{3}\right)^{-1}\right)^{1 / 3}\left(1+N^{-1 / 2} l^{-3} u^{3}\right)^{1 / 3} \frac{l^{3} N^{1 / 2}}{u} d \Omega_{7}^{2},\\
C^{ t x y}& =-\frac{1}{8} \left(1+\rho_{\dot{\alpha}} x^{\dot{\alpha}} \left(1+N^{-1 / 2} l^{-3} u^{3}\right)^{-1}\right) \left(1+N^{-1 / 2} l^{-3} u^{3}\right)^{-1}.
\end{aligned}
\end{equation}
{Now,} fixing the deformation parameter $\hat{\rho}_{\alpha_{1}}=N^{1 / 2} {\rho}_{\alpha_{1}}$ in the near-horizon, we get the following background:\vspace{-6pt}
\begin{equation} 
\begin{aligned}
d s_{(h)}^{2}=&\ \frac{l^{2}}{4}\left(u^{3}+l^{3} \hat{\rho}_{\dot{\alpha}} x^{\dot{\alpha}}\right)^{-2 / 3}\left(-d t^{2}+d x^{2}+d y^{2}\right)\\
&+\frac{1}{u}\left(u^{3}+l^{3} \hat{\rho}_{\dot{\alpha}} x^{\dot{\alpha}}\right)^{1 / 3}\left(\frac{l^{2}}{4 u^{2}} d u^{2}+d \Omega_{7}^{2}\right), \\
C_{(h) t x y}=&-\frac{1}{8}\left(1+\hat{\rho}_{\dot{\alpha}} x^{\dot{\alpha}} \frac{l^{3}}{u^{3}}\right)^{-1}.
\end{aligned}
\end{equation}
{This} is exactly the $P \wedge P \wedge M$ deformation of the AdS${}_4 \times\SS^7$ solution of \cite{Bakhmatov:2020kul}. Note that this deformation is non-unimodular in the full space-time of the M2-brane, including the near-horizon area. Hence, we conclude that both unimodular and non-unimodular  tri-vector deformations commute with the near-horizon limit.

This result shows that there are no tri-vector deformations of the  AdS$_4\times \SS^7$ background preserving SUSY as well, at least in the SL(5) setup. Indeed, since deformation and using the near-horizon limit commute is the only way to  preserve SUSY, we get to keep some of the supersymmetry restored in the limit, which is an 
additional 16 spinors. However, the second equation in  system \eqref{eq:system} does not change when the limit is taken, and  $\det \G^4 \neq 0$ renders $\e=0$. Explicit calculation using AdS metric for given deformations gives the same result.

\section{Conclusions}\label{sec5}

In this work, we consider conditions under which a tri-vector deformation given by an SL(5) transformation parametrized by $W_m=1/3! \e_{mnkl}\W^{nkl}$ preserves the supersymmetry of 11D backgrounds. Our results give a particular generalization of those presented in~\cite{Orlando:2018qaq} for bi-vector deformations preserving the supersymmetry of 10D backgrounds. The main idea behind our approach is to notice that the SL(5) tri-vector deformation breaks the upper-triangular parametrization of a generalized vielbein defining the supergravity frame of the SL(5) exceptional field theory. To restore it, one performs an additional USp(4) $<$ SL(5) transformation $K$, which depends on the deformation parameter $W_m$ and background fields. This local transformation acts on indices of fermionic fields as well as on the Killing spinor entering BPS equations. Requiring the BPS equations to hold, we arrive at the desired condition \eqref{eq:susycondzero}.

To write BPS equations for the fields of the SL(5) ExFT, we first derive supersymmetry transformations of the theory. This we perform by first imposing them in a general form inspired by the E$_{6(6)}$ supersymmetric ExFT of \cite{Musaev:2014lna} and then requiring them to satisfy the correct algebra of local symmetries of the theory and to reproduce SUSY rules of maximal $D~=~7$ gauged supergravity. This fixes all free coefficients in transformations up to a single one, which gets absorbed into a single field redefinition. 

The general setup of the tri-vector deformation formalism within the SL(5) theory, as defined in \cite{Bakhmatov:2020kul}, significantly restricts the number of possible examples to check against the general formula. In particular, only backgrounds of the form $M_4\times M_7$ with three forms  in the directions of $M_4$ are allowed
. Given that we investigate supersymmetry preservation under deformations of the M2-brane background and of AdS${}_4\times \SS^7$ as its near-horizon limit, the result is negative: no deformation within the setup preserves any supersymmetry. This provides a few directions in which the research can be continued.

The most interesting and suggestive would be to construct a poly-vector deformation scheme for a full E$_{d(d)}$ theory (SL(5) for $d=4$), extending the results of \cite{Bakhmatov:2020kul,Gubarev:2020ydf} to backgrounds with non-diagonal components in the full 11D metric and a more general 3-form field. On the one hand, this could change  condition \eqref{eq:susycondzero}; on the other hand, this would allow us to consider more general examples of deformed backgrounds and hopefully find ones with preserved supersymmetries. Another approach that would extend the space of possibilities is to allow non-unimodular deformations, i.e., $\nabla_{[m}W_{n]}\neq 0$. This will, in general, move us out of the space of supergravity solutions generating backgrounds to solve equations of the generalized 11D supergravity of \cite{Bakhmatov:2022rjn,Bakhmatov:2022lin}. This is an 11-dimensional uplift of the 10-dimensional generalized supergravity \cite{Arutyunov:2015mqj}.

\section*{Acknowledgments}
This work  has been supported by Russian Science Foundation grant RSCF-20-72-10144 and in part by the Foundation for the Advancement of Theoretical
Physics and Mathematics “BASIS”, grant No 21-1-2-3-1. The work of EtM has been partially funded by Russian Ministry of Education and Science.

\appendix

\section{Conventions and Notations}
\label{sec:proj}

Generators in the $\bf5$ and $\bf 10$
\begin{equation}
    \begin{aligned}
        (t_M{}^N)_K{}^L &= \delta_M{}^L \delta_K{}^N - \fr15 \delta_M{}^N \delta_K{}^L,\\
        (t_M{}^N)_{KL}{}^{PQ} & = 4 (t_M{}^N)_{[K}{}^{[P}\delta_{Q]}{}^{L]}.
    \end{aligned}
\end{equation}
{The} factor in the second line has been chosen such that the commutation relations read
\begin{equation}
    [t_M{}^N, t_K{}^L] = \delta_K{}^N t_M{}^L - \delta_M{}^L t_K{}^N
\end{equation}
and contraction of indices $\cM,\cN$ labelling  $\bf 10$ is performed by an additional prefactor \mbox{of 1/2:}
\begin{equation}
    A^\cM B_{\cM } = \fr12 A^{MN}B_{MN}.
\end{equation}

Projectors on the adjoint representation of SL(5) in  $\bf 5$ and in the mixed representation are given by
\begin{equation}
    \begin{aligned}
        \PP^{M}{}_{N}{}^{K}{}_{L}& = (t_{Q}{}^{P})_{N}{}^{M}(t_{P}{}^{Q})_{L}{}^{K},\\
        \PP^{M}{}_{N}{}^{K L}{}_{P Q} & = \PP^{M}{}_{N}{}^{[K}{}_{[P} \delta_{Q]}{}^{L]}.
    \end{aligned}
\end{equation}
{These} satisfy
\begin{equation}
    \begin{aligned}
        \PP^{M}{}_{N}{}^{K}{}_{L}\PP^{L}{}_{K}{}^{P}{}_{Q} & = \PP^{M}{}_{N}{}^{P}{}_{Q} ,\\
        \PP^{M}{}_{N}{}^{N}{}_{M} & = \mbox{dim}(adj) = 24,\\
        \fr14\PP^{M}{}_{N}{}^{P Q}{}_{R S}\PP^{K}{}_{L}{}^{R S}{}_{ P Q} & = 3 \PP^{M}{}_{N}{}^{K}{}_{L}.
    \end{aligned}
\end{equation}

Some useful gamma-matrices relations:
\begin{equation}
    \begin{aligned}
        \g^{\m\rho\s} &= \g^{\rho\s}\g^\m + 2 g^{\m[\rho}\g^{\s]},\\
        \g^{\m\rho\s} &= \g^\m\g^{\rho\s} - 2 g^{\m[\rho}\g^{\s]}
    \end{aligned}
\end{equation}
\begin{equation}
[\gamma_{\mu},\gamma^{\rho \lambda}]=2\delta^{\rho}_{\mu} \gamma^{\lambda}-2 \delta^{\lambda}_{\mu} \gamma^{\rho}
\end{equation}
\begin{equation}
\gamma^{\mu_5} \gamma^{\mu_1 \mu_2 \mu_3 \mu_4} =\gamma^{\mu_1 \mu_2 \mu_3 \mu_4 \mu_5} +4g^{\mu_5 [\mu_1}  \gamma^{ \mu_2 \mu_3 \mu_4]}
\end{equation}
\begin{equation}
\gamma^{\mu_1 \mu_2 \mu_3} =\gamma^{\mu_3}\gamma^{\mu_1 \mu_2}+2\gamma^{[\mu_1}g^{\mu_2] \mu_3} = \gamma^{\mu_1 \mu_2}\gamma^{\mu_3}-2\gamma^{[\mu_1}g^{\mu_2] \mu_3}
\end{equation}

Using the USp(4) invariant tensor $\Omega_{AB}$, it is possible to define an analog of the epsilon-tensor of SL(4), which defines relations between, say, two realizations of the $\bf 4$ irreducible representation, $T_{[ABC]}$ and $T^A$:
\begin{equation}
    \W_{ABCD} = 3\W_{[AB}\W_{CD]}.
\end{equation}
{The} prefactor is chosen in order to ensure that $\Omega_{ABCD}$ has the same properties under contraction with $\Omega^{ABCD}$ as the epsilon-tensor.

Sometimes we use the following rewriting of fields in  $\bf 10$:
\begin{equation}
    \W_{CD}T_{[MN]}\mV^{M AC}\mV^{NBD} = \fr{\sqrt{2}}{4}T^{(AB)}.
\end{equation}

\section{Supersymmetry Rules}

\label{app:susy}

Here we perform all necessary checks for supersymmetry transformations of the SL(5) exceptional field theory. We start with transformations of the gravitino and dilatino fields
\begin{equation}
\begin{aligned}
\delta \psi_{\mu}^{A}=&\ \mathcal{D}_{\mu} \epsilon^{A}+\Omega_{CD } \mathcal{V}^{M AC} \mathcal{V}^{N BD} \Omega_{BF}\left(\alpha_{11} \nabla_{M N}^{+}\left(\gamma_{\mu} \epsilon^{F}\right)+\alpha_{12} \gamma_{\mu} \nabla_{M N}^{+} \epsilon^{F}\right)\\
&+\alpha_{13} \mathcal{V}_{BC}^{N} \mF_{\nu \rho \lambda N} \Omega^{AB}\left(\gamma^{\nu \rho \lambda}{ }_{\mu}+\frac{9}{2} \gamma^{\nu \rho} \delta_{\mu}^{\lambda}\right) \epsilon^{C},\\
\delta \chi^{ABC}=&\ 2 \Omega^{CD} \mathcal{P}_{\mu DE}{}^{AB} \gamma^{\mu} \epsilon^{E}+\alpha_{21} \Omega_{ED} \mathcal{V}^{M CE} \mathcal{V}^{N [A  \mid D \mid} \nabla_{M N}^{+} \epsilon^{B]}\\
&+\alpha_{22}  \left(\Omega^{AB} \delta_{G}^{C}-\Omega^{C[A} \delta_{G}^{B]}\right) \Omega_{DE} \Omega_{FH} \mathcal{V}^{M G F} \mathcal{V}^{N D H} \nabla_{M N}^{+} \epsilon^{E}  \\
&+\alpha_{23}\bigg(\Omega^{AD} \Omega^{BE} \mathcal{V}_{DE}{ }^{L} \mF_{\mu \nu \rho L} \gamma^{\mu \nu \rho} \epsilon^{C}\\
&-\frac{1}{5}\left(\Omega^{AB} \Omega^{CF}+4 \Omega^{C[A} \Omega^{B] F}\right) \mathcal{V}_{FE}{ }^{L} \mF_{\mu \nu \rho L} \gamma^{\mu \nu \rho} \epsilon^{E}\bigg)
\end{aligned}
\end{equation}
here
\begin{equation}
    \begin{aligned}
        \nabla_{M N}^{\pm}\epsilon^{A}=&\  \partial_{M N}\epsilon^{A}+ \frac{1}{4}e^{\mu \a} \partial_{M N}e_{\mu}^{\beta} \gamma_{\a\beta}\epsilon^{A}   \pm \alpha_{0} \frac{1}{4} \mF_{\mu \nu M N} e^{\mu \a} e^{\nu \beta} \gamma_{\a\beta}\epsilon^{A}\\
        &-\cQ_{MN}{}_B{}^A \epsilon^{B}   + \fr53 \lambda_\e \G_{K[M,N]}{}^K \e^A.      
     \end{aligned}
\end{equation}
Comparing to the SUSY transformation rules of the ungauged $D~=~7$ supergravity in the notations of \cite{Samtleben:2005bp}, we have
\begin{equation}
    \begin{aligned}
        \a_{12} & = -\fr35 \a_{11}, && \a_{22} = -\fr15 \a_{21}, && \a_{21} = 10 \a_{11},\\
        \a_{23} & = \fr53 \a_{13},
    \end{aligned}
\end{equation}
which leaves three coefficients. These can be determined by fixing the relations between the field strengths $F$ of ExFT and $\mH$ of $D~=~7$ maximal supergravity, and between the coset fields and components of the generalized metric, which we perform later. 

Supersymmetry transformations for bosonic fields can be composed in the following~form:
\begin{equation}
    \begin{aligned}
        \delta e_{\mu}^{\a}&=\frac{1}{2} \bar{\epsilon}_{A} \gamma^{\a} \psi_{\mu}^{A},\\
        \beta_1 \delta A_{\mu}^{M N}&=-V_{a b}^{[M } \mathcal{V}_{c d}^{N]} \Omega^{b d}\left(\frac{1}{2} \Omega^{a e} \bar{\epsilon}_{e} \psi_{\mu}^{c}+\frac{1}{4} \bar{\epsilon}_{e} \gamma_{\mu} \chi^{c a e}\right),\\
        \delta \mathcal{V}_{M}^{\mathrm{ab}}&=\frac{1}{4} \mathcal{V}_{M}^{\mathrm{cd}}\left(\Omega_{e[c} \bar{\epsilon}_{d]}  \chi^{abe}+\frac{1}{4} \Omega_{c d} \bar{\epsilon}_{e} \chi^{a b e}+\Omega_{c e} \Omega_{d f} \bar{\epsilon}_{g} \chi^{c f[a} \Omega^{b] g}+\frac{1}{4} \Omega_{c e} \Omega_{d f} \Omega^{a b} \bar{\epsilon}_{g} \chi^{c f g}\right),\\
        \beta_2 \delta B_{\mu \nu M}&=\mathcal{V}_{M}^{a b}\left(-\Omega_{a c} \bar{\epsilon}_{b} \gamma_{[\mu} \psi_{\nu]}^{c}+\frac{1}{8} \Omega_{a c} \Omega_{b d} \bar{\epsilon}_{e} \gamma_{\mu \nu} \chi^{c d e}\right)+2 {\beta_1}^2\epsilon_{M N P Q R} A_{[\mu}^{N P} \delta A_{\nu \mid}^{Q R},\\
        \Delta C_{\mu \nu \rho}^{M}&=\mathcal{V}_{a b}^{M}\left(-\frac{3}{8} \Omega^{a c} \bar{\epsilon}_{c} \gamma_{[\mu \nu} \psi_{\rho]}^{b}-\frac{1}{32} \bar{\epsilon}_{c} \gamma_{\mu \nu \rho} \chi^{a b c}\right).
    \end{aligned}
\end{equation}
{Transformation} of the 3-form field
\begin{equation}
 \Delta C_{\mu \nu \rho}^{N} \equiv \left( \beta_3  \delta C_{\mu \nu \rho}^{N}-3 \beta_1 \beta_2 B_{[\mu \nu P} \delta A_{\rho]}^{P N}+2{\beta_1}^3 \epsilon_{P Q R S T} A_{[\mu}^{N P} A_{\nu}^{Q R} \delta A_{\rho]}^{S T}\right),
\end{equation}
fixes the coefficient $\beta_3$.

Let us now check that the supersymmetry transformation, say, on the vielbein,  closes 
correctly into the algebra of symmetries of the theory. For that, we consider commutator
\begin{equation}
\begin{aligned}[]
[\delta_{\epsilon_1},\delta_{\epsilon_2}]  e_{\mu}{}^{\alpha}  =&\ \frac{1}{2} \bar{\epsilon}_{ 2 A} \gamma^{\alpha} \delta_{\epsilon_1} \psi_{\mu}^{A}  -  (1 \leftrightarrow 2) = \frac{1}{2} \bar{\epsilon}_{ 2 A} \gamma^{\alpha} \mathcal{D}_{\mu} \epsilon^{A}_1\\
 &+\Omega_{C D } \Omega_{BF} \frac{1}{2}  \bar{\epsilon}_{ 2 A} \gamma^{\alpha} \mathcal{V}^{M AC} \mathcal{V}^{N BD}\left(\alpha_{11} \nabla_{M N}^{+}\left(\gamma_{\mu} \epsilon^F_{1}\right)+\alpha_{12}  \gamma_{\mu} \nabla_{M N}^{+} \epsilon^F_{1}\right) \\
& +\alpha_{13} \frac{1}{2}  \bar{\epsilon}_{ 2 A} \gamma^{\alpha} \mathcal{V}_{B C}^{N} F_{\nu \rho \lambda N} \Omega^{AB}\left(\gamma^{\nu \rho \lambda}{ }_{\mu}+\frac{9}{2} \gamma^{\nu \rho} \delta_{\mu}^{\lambda}\right) \epsilon^{C}_1 -  (1 \leftrightarrow 2)
\end{aligned}
\end{equation}
{Using} $\gamma$-matrices relations, we rewrite the terms with derivative $\mc{D}_\m$ hitting the SUSY parameter:
\begin{equation}
\frac{1}{2} \bar{\epsilon}_{ 2 A} \gamma^{\alpha} \mathcal{D}_{\mu} \epsilon^{A}_1 - \frac{1}{2} \bar{\epsilon}_{ 1 A} \gamma^{\alpha} \mathcal{D}_{\mu} \epsilon^{A}_2 = \frac{1}{2} \bar{\epsilon}_{ 2 A} \gamma^{\alpha} \mathcal{D}_{\mu} \epsilon^{A}_1 + \frac{1}{2} \mathcal{D}_{\mu} \bar{\epsilon}_{2A} \gamma^{\alpha}  \epsilon^{A}_1 =
\frac{1}{2} \mathcal{D}_{\mu} ( \bar{\epsilon}_{2A} \gamma^{\alpha}  \epsilon^{A}_1)
\end{equation}
{For} the terms containing $\nabla_{MN}^+$, we have the following:
\begin{equation}
\begin{aligned}
& \Omega_{CD} \Omega_{BF} \frac{1}{2} \mathcal{V}^{M AC} \mathcal{V}^{N BD} \alpha_{11}   (\bar{\epsilon}_{ 2 A} \gamma^{m} \nabla_{M N}^{+}\left(\gamma_{\mu} \epsilon_{1}^F\right)- \bar{\epsilon}_{ 1 A}\gamma^{\alpha}  \nabla_{M N}^{+}\left(\gamma_{\mu} \epsilon_{2}^F \right) ) \\
=&\ \Omega_{CD } \Omega_{BF} \frac{1}{2} \mathcal{V}^{M AC} \mathcal{V}^{N BD} \alpha_{11}   \left(\bar{\epsilon}_{ 2 A}\left( \gamma^{\alpha}{}_{\beta} +\delta^{\alpha}_{\beta} \right) \epsilon_{1}^F   \nabla_{M N}^{+} e^{\beta}_{\mu}  + \bar{\epsilon}_{ 2 A}\left( \gamma^{\alpha}{}_{\beta} +\delta^{\alpha}_{\beta} \right)  e^{\beta}_{\mu}  \nabla_{M N}^{+} \epsilon_{1}^F \right. \\
& \left. - \bar{\epsilon}_{ 1 A} \left( \gamma^{\alpha}{}_{\beta} +\delta^{\alpha}_{\beta} \right)  \nabla_{M N}^{+} \epsilon_{2}^F  e^{\beta}_{\mu} - \bar{\epsilon}_{ 1 A} \left( \gamma^{\alpha}{}_{\beta} +\delta^{\alpha}_{\beta} \right)  \epsilon_{2}^F \nabla_{M N}^{+} e^{\alpha}_{\mu}    \right) \\
 =&\ \Omega_{CD } \Omega_{BF} \frac{1}{2} \mathcal{V}^{M AC} \mathcal{V}^{N BD} \alpha_{11}   \Big(  2\delta^{\alpha}_{\beta} \bar{\epsilon}_{ 2 A} \epsilon_{1}^F   \nabla_{M N}^{+} e_{\mu}{}^{\beta} + \delta^{\alpha}_{\beta} \nabla_{M N}^{+} \left(\bar{\epsilon}_{ 2 A} \epsilon_{1}^F\right) e_{\mu}{}^{\beta} \\
&+ \bar{\epsilon}_{ 2 A} \gamma^{m}{}_{n} \nabla_{M N}^{+} \epsilon_{1}^F e^n_{\mu}- \nabla_{M N}^{+} \bar{\epsilon}_{ 2 A} \gamma^{\alpha}{}_{\beta}  \epsilon_{1}^F e_{\mu}{}^{\beta} \Big)
\end{aligned}
\end{equation}
{This} can be simplified as follows:
\begin{equation}
\begin{aligned}
\Omega_{CD} \Omega_{BF} \frac{1}{2}   \mathcal{V}^{M AC} \mathcal{V}^{N BD}  \alpha_{12} \left( \bar{\epsilon}_{ 2 A}  \left(\gamma^{\alpha}{}_{\mu} +e_{\mu}{}^{\alpha} \right) \nabla_{M N}^{+} \epsilon^F_{1} - \bar{\epsilon}_{ 1 A} \left( \gamma^{\alpha} {}_{\mu} + e_{\mu}{}^{\alpha}  \right)\nabla_{M N}^{+} \epsilon^F_{2} \right) \\
=\Omega_{CD } \Omega_{BF} \frac{1}{2}   \mathcal{V}^{M AC} \mathcal{V}^{N BD}  \alpha_{12} \left( e^{m}_{\mu} \nabla_{M N}^{+} \left(\bar{\epsilon}_{ 2 A} \epsilon_{1}^F\right)  + \bar{\epsilon}_{ 2 A} \gamma^{m}{}_{\mu} \nabla_{M N}^{+} \epsilon_{1}^F   - \nabla_{M N}^{+} \bar{\epsilon}_{ 2 A} \gamma^{m}{}_{\mu}  \epsilon_{1}^F   \right)
\end{aligned}
\end{equation}
{Finally}, we rewrite the remaining terms as
\begin{equation}
\begin{aligned}
&\alpha_{13} \frac{1}{2}  \bar{\epsilon}_{ 2 A} \gamma^{\alpha} \mathcal{V}^N{}_{BC} F_{\nu \rho \lambda N} \Omega^{AB}\left(\gamma^{\nu \rho \lambda}{ }_{\mu}+\frac{9}{2} \gamma^{\nu \rho} \delta_{\mu}^{\lambda}\right) \epsilon^{C}_1 \\
&- \alpha_{13} \frac{1}{2}  \bar{\epsilon}_{ 1 A} \gamma^{\alpha} \mathcal{V}^N{}_{BC} F_{\nu \rho \lambda N} \Omega^{A B}\left(\gamma^{\nu \rho \lambda}{ }_{\mu}+\frac{9}{2} \gamma^{\nu \rho} \delta_{\mu}^{\lambda}\right) \epsilon^{C}_2 \\
&=\alpha_{13} \frac{1}{2}\mathcal{V}^N{}_{BC} F_{\nu \rho \lambda N} \Omega^{AB}\left(\bar{\epsilon}_{ 2 A} \left(\gamma^{\nu \rho \lambda}{}_{\mu}{}^{ \alpha} +4g^{\alpha [\nu}  \gamma^{ \rho \lambda]}{}_{\mu} \right)\epsilon^{C}_1+\frac{9}{2}\bar{\epsilon}_{ 2 A} \left( \gamma^{\nu \rho \alpha}-2\gamma^{[\nu}g^{\rho] \alpha} \right) \delta_{\mu}^{\lambda}\epsilon^{C}_1 \right)  \\
&-  (1 \leftrightarrow 2)=\alpha_{13} \mathcal{V}^{N}{}_{BC} F_{\nu \rho \lambda N} \Omega^{AB}\bar{\epsilon}_{ 2 A} \left(  \gamma^{\nu \rho \lambda}{}_{\mu}{}^{ \alpha}+9 g^{\alpha \nu}\gamma^{\rho} \delta_{\mu}^{\lambda}  \right)  \epsilon^{C}_1 
\end{aligned}
\end{equation}
{Collecting} the above together, for the commutator we obtain:
\begin{equation}
\begin{aligned}[]
[\delta_{\epsilon_1},\delta_{\epsilon_2}] e_{\mu}{}^{\alpha}&= \frac{1}{2} \mathcal{D}_{\mu} ( \bar{\epsilon}_{2A} \gamma^{\nu}  \epsilon^{A}_1 e_{\nu}^{\alpha}) + \alpha_{11} \Omega_{CD } \Omega_{BF}  \mathcal{V}^{M AC} \mathcal{V}^{N BD}  \bar{\epsilon}_{ 2 A} \epsilon_{1}^F   \nabla_{M N}^{+} e_{\mu}{}^{\alpha} \\
&+(\alpha_{11}+\alpha_{12}) \Omega_{CD } \Omega_{BF} \frac{1}{2}   \mathcal{V}^{M AC} \mathcal{V}^{N BD}    \nabla_{M N}^{+} \left(\bar{\epsilon}_{ 2 A} \epsilon_{1}^f\right)e^{m}_{\mu}   \\
&+\left((\alpha_{11}+\alpha_{12}) \Omega_{CD } \Omega_{BF} \frac{1}{2}   \mathcal{V}^{M AC} \mathcal{V}^{N BD} \left (\bar{\epsilon}_{ 2 A} \gamma^{{}^{\alpha}}{}_{\beta} \nabla_{M N}^{+} \epsilon_{1}^f   - \nabla_{M N}^{+} \bar{\epsilon}_{ 2 A} \gamma^{\alpha}{}_{\beta}  \epsilon_{1}^F   \right) \right.  \\
&\left.+ \, \alpha_{13} \mathcal{V}^N{}_{BC} F_{\nu \rho \lambda N} \Omega^{AB}\bar{\epsilon}_{ 2 A} \left(  \gamma^{\nu \rho \lambda}{}_{\beta}{}^{\alpha}+9 g^{m \nu}\gamma^{\rho} \delta_{\beta}^{\lambda}  \right)  \epsilon^{C}_1  \right) e_{\mu}{}^{\alpha}
\end{aligned}
\end{equation}
{In} the final step, we use the generalized vielbein postulate and vanishing torsion condition to rewrite the commutator in the following form
\begin{equation}
[\delta_{\epsilon_1},\delta_{\epsilon_2}] e_{\mu}{}^{\alpha} =\mathcal{D}_{\mu} \xi^{\nu} e_{\nu}{}^{\alpha}+ \xi^{\nu}\mathcal{D}_{\nu}e^n_{\mu} +\Lambda^{MN} \partial_{MN} e_{\mu}{}^{\alpha}+
\frac{(\alpha_{11}+\alpha_{12})}{2\alpha_{11}}\Lambda^{MN} \partial_{MN}\Lambda^{MN}+\Lambda^{\alpha}{}_{\beta} e_{\mu}{}^{\beta}.
\end{equation}
{Here,} on the RHS, we recognize external diffeomorphisms, generalized Lier derivatives,  and the $\so(1,6)$ Lorentz transformation with parameters given by 
\begin{equation}
    \begin{aligned}
        \xi^{\nu}&=\bar{\epsilon}_{2A} \gamma^{\nu}  \epsilon^{A}_1, \\
        \Lambda^{MN} &=  \alpha_{11} \Omega_{CD } \Omega_{BF}  \mathcal{V}^{M AC} \mathcal{V}^{N BD}  \bar{\epsilon}_{ 2 A} \epsilon_{1}^F, \\
        \Lambda^{\alpha}{}_{\beta}&=\left((\alpha_{11}+\alpha_{12}) \Omega_{CD } \Omega_{BF} \frac{1}{2}   \mathcal{V}^{M AC} \mathcal{V}^{N BD} \left (\bar{\epsilon}_{ 2 A} \gamma^{\alpha}{}_{{\beta}} \nabla_{M N}^{+} \epsilon_{1}^f   - \nabla_{M N}^{+} \bar{\epsilon}_{ 2 A} \gamma^{m}{}_{n}  \epsilon_{1}^F   \right) \right. \\
        &\left.+ \, \alpha_{13} \mathcal{V}^N{}_{BC} F_{\nu \rho \lambda N} \Omega^{AB}\bar{\epsilon}_{ 2 A} \left(  \gamma^{\nu \rho \lambda}{}_{\beta}{}^{ \alpha}+9 g^{\alpha \nu}\gamma^{\rho} \delta_{\beta}^{\lambda}  \right)  \epsilon^{C}_1  \right)-\Lambda^{MN}\omega^{+}_{MN}{}^{\alpha}{}_{\beta}
\end{aligned}
\end{equation}
{Similarly,} the closure of the algebra can be checked for all other fields, which we prefer not to go through here. Indeed, all structures of ExFT have  already been used in the calculation above and one based on the similar calculation for the E${}_{6(6)}$ case \cite{Musaev:2014lna}, we do not expect new issues to come up but simply various fixes of arbitrary coefficients. Instead, we perform a reduction to the maximal $D~=~7$ ungauged supergravity, which is already enough to fix the transformations.

First, comparing the Lagrangian of the maximal $D~=~7$ SUGRA and the SL(5) ExFT, we fix coefficients in the bosonic supersymmetry rules
\begin{equation}
\beta_1= \frac{1}{2\sqrt{2}}, \, \beta_2=\frac{1}{8\sqrt{2}}, \, \beta_3=-\frac{1}{32}.
\end{equation}
{Hence}, these are simply due to various field rescalings. Next, we compare SUSY rules of ExFT when $\partial_{MN}=0$ to those of the maximal $D~=~7$ SUGRA, which gives
\begin{equation}
    \begin{aligned}
        \alpha_{12} &= -\frac{3}{5} \alpha_{11},  &&
        \alpha_{22}=-\frac{1}{5} \alpha_{21} \\
        \alpha_0 \alpha_{11} & =-\frac{4}{5}, &&
        \alpha_{13} =-\frac{1}{15}, \\
        \alpha_{0}\alpha_{21}&=-8, &&
        \alpha_{23}=-\frac{1}{6}.
    \end{aligned}
\end{equation}
{This} leaves one free coefficient $\a_0$ that can be reabsorbed into the remaining field redefinitions and we set $\a_0=1$. This completely determines the SUSY rules.

\bibliography{bib.bib}
\bibliographystyle{utphys.bst}

\end{document}

%% file: Defs.tex
\def\a{\alpha} 
 
\def\g{\gamma} 
 
\def\e{\epsilon}
\def\ve{\varepsilon} 
 
\def\h{\eta}

\def\m{\mu}
\def\n{\nu}

\def\q{\theta}
\def\s{\sigma} 
  
\def\f{\phi}
\def\y{\psi} 
 
\def\w{\omega}

\def\G{\Gamma} 
\def\D{\Delta}

\def\P{\Pi}

\def\W{\Omega}

\def\fr{\frac} \def\dfr{\dfrac} \def\dt{\partial}

\def\mT{\mathcal{T}}
\def\mc{\mathcal}

\def\mD{\mathcal{D}}

\def\mF{\mathcal{F}}

\def\mH{\mathcal{H}}
\def\mL{\mathcal{L}}
\def\mH{\mathcal{H}}

\def\mQ{\mathcal{Q}}

\def\mV{\mathcal{V}}

\def\PP{\mathbb{P}}

\def\DD{{\mathcal{D}}}

\def\cM{{\mathcal{M}}}
\def\cN{{\mathcal{N}}}

\def\cQ{{\mathcal{Q}}}

\def\cV{{\mathcal{V}}}

\def\XX{\mathbb{X}}
\def\RR{\mathbb{R}}
\def\SS{\mathbb{S}}

\def\bas{\mathrm{bas}\,}

\def\sl{{\mathfrak{sl}}}
\def\gl{{\mathfrak{gl}}}

\def\so{{\mathfrak{so}}}

\def\SO{\mathrm{SO}}

\def\SL{\mathrm{SL}}

\def\USp{\mathrm{USp}}

\def\diag{\mathrm{diag}}